\documentclass[5p,number,sort&compress]{elsarticle}

\usepackage{vmargin}
\usepackage{graphicx}
\usepackage{subcaption}
\usepackage{hyperref}
\usepackage{amsmath}
\usepackage{natbib}
\usepackage{pifont}
\usepackage{geometry}
\usepackage{fleqn}
\usepackage{txfonts}

\usepackage{xcolor}

\usepackage{lineno}

\begin{document}


\title{Neutron-$\gamma$ discrimination with organic scintillators: Intrinsic pulse shape and light yield contributions}

\author[]{M. S\'enoville \fnref{fn1}}
\fntext[fn1]{Present address: Neuro-PSI, CNRS, Universit\'e Paris-Sud, Gif-Sur-Yvette, France}
\author[]{F. Delaunay\corref{corr}} \ead{delaunay@lpccaen.in2p3.fr}
\cortext[corr]{Corresponding author}
\author[]{M. P\^arlog}
\author[]{N. L. Achouri}
\author[]{N. A. Orr}
\address{LPC Caen, Normandie Universit\'e, ENSICAEN, UNICAEN, CNRS/IN2P3, Caen, France}

\begin{abstract}
A comparative study of the neutron-$\gamma$ Pulse Shape Discrimination (PSD) with seven organic scintillators is performed using an identical setup and digital electronics.
The scintillators include plastics (EJ-299-33 and a plastic prototype), single crystals (stilbene and the recent doped $p$-terphenyl) and liquids (BC501A, NE213 and the deuterated liquid BC537).
First, the overall PSD performance of the different scintillators is compared and threshold neutron energies for a given discrimination quality are determined. 
Then, using statistical arguments, two intrinsic contributions to the PSD capability of the scintillating materials are disentangled: the light yield and the specific pulse shapes induced by neutrons and $\gamma$-rays.
This separation provides additional insight into the behaviour of organic scintillators and allows a detailed comparison of the discrimination performance of the various materials.
On the basis of this analysis, limitations of current organic scintillators and of recently proposed alternative scintillators are discussed.

\end{abstract}


\maketitle


\section{Introduction}

Organic scintillators are particularly well suited to neutron detection:
they are rich in hydrogen and usable in large volumes (\textit{e.g.} \cite{ORION, BNB, CARMEN}); they give high detection efficiencies; the short time constant (typically a few ns) of their prompt scintillation allows a good time resolution beneficial for neutron time-of-flight spectroscopy (\textit{e.g.} \cite{EDEN, DEMON, NWALL, TONNERRE});
they can be loaded with isotopes offering large neutron cross-sections \cite{ORION, BNB, CARMEN, Normand, Bass}.
In addition, as discussed below, various organic scintillators allow pulse shape discrimination.

The scintillation of organic materials corresponds to radiative transitions from the first singlet excited states $S_1$ to the ground states $S_0$ of aromatic $\pi$-electron systems.
Detailed presentations of organic scintillation can be found in text books \cite{Birks64,Knoll} and review articles \cite{BrooksReviewNIM}.
Here we recall the main aspects as a basis for the present discussions.

The prompt scintillation originates from $S_1$ states populated by fast ($10^{-11}$ to $10^{-10}$ s) non-radiative transitions from high-lying singlet states $S_n$ directly excited by a charged particle.
However, interactions with other excited or ionized molecules can induce the complete dissipation of the $S_n$ energy, resulting in a loss of scintillation. The effect of this so-called ``ionization quenching'' \cite{Birks64,BrooksReviewNIM} increases with the stopping power of the charged particle.
Triplet $T_1$ states at the origin of the delayed scintillation are produced mainly by non-radiative transitions from highly excited $T_n$ states populated by ionization and recombination.
The recombination time ($\approx 10^{-10}$ s) makes this path less sensitive to ionization quenching.
As $T_1 \rightarrow S_0$ transitions are inhibited by the multiplicity selection rule, $T_1$ states deexcite via $T_1 + T_1 \rightarrow S_1 + S_0$ Triplet-Triplet Annihilation (TTA) followed by $S_1$ decay.
This leads to the delayed scintillation emission with a typical lifetime of a few hundred ns \cite{BrooksReviewNIM}.
Since TTA is bimolecular, its rate increases with the density of $T_1$ states, \textit{i.e.} with the stopping power of the charged particle.

The above processes are widely accepted as the origin of the increase of the delayed light fraction with the stopping power observed in some organic scintillators \cite{BrooksReviewNIM}.
This results in the dependence of the scintillation intensity time profile (``pulse shape'') on the nature of the exciting particle, thus allowing particle identification through pulse shape discrimination (PSD) techniques.
Classical examples are the discrimination of neutrons and $\gamma$-rays, or of $\alpha$-particles and $\gamma$-rays \cite{Brooks1959}.
This PSD capability is crucial in particular for the detection of fast neutrons in a $\gamma$-ray background.

Until recently, the only organic scintillators featuring this discrimination capability were crystals such as anthracene, stilbene, and liquid scintillators, among which NE213 and its equivalents BC501A and EJ301 have been the most widely used. Some deuterated scintillators such as the equivalent liquids NE230, BC537 and EJ315 also present a PSD capability.
Unfortunately the use of these materials has been limited by some of their properties:
crystals show an anisotropic response \cite{Brooks1974} and their dimensions are limited to some 10 cm \cite{Budakovsky}; liquids present an intrinsic spill risk and many of them are toxic and flammable with a low flash point. As such they have been excluded from many facilities and applications.

It was shown in 1960 that plastic scintillators can also present PSD capabilities comparable to those of crystals or liquids \cite{Brooks60}, although most of them offer poor discrimination \cite{Winyard}.
However, it is only a few years ago that plastic scintillators offering good PSD became commercially available. Following new developments \cite{Zaitseva}, Eljen Technology has produced and commercialised the EJ299-33 and EJ299-34 plastics, very recently replaced by the improved EJ276 plastic \cite{Eljen,Zaitseva2018}.
Without the drawbacks of the liquids, these discriminating plastics will find larger applicability.

A plastic scintillator is a solution of one or two scintillating compounds (``solutes'') in a polymer matrix.
One key factor to obtain good PSD with plastics is to increase the TTA probability between two solute molecules. This is achieved with high solute concentrations \cite{Zaitseva}, but this leads to mechanical softness and limited plastic lifetime.
Recent developments have focused on solving such issues by modifications in the plastic formulations \cite{Zaitseva2018,Blanc,Zhmurin}.

Neutron-$\gamma$ PSD with organic scintillators has been explored extensively, including new discriminating plastics \cite{Zaitseva2018,PozziEJ299,NyibuleEJ299,CesterEJ299,LawrenceEJ299}.
However, most of the studies are restricted to a pair of scintillators, or to scintillators of the same type, and comparing results from different works is delicate due to various experimental setups and conditions, or different choices for the evaluation of the PSD quality. Furthermore, the origins of the differences in PSD performance of the scintillators are usually not or partially discussed.

In this work we characterise and compare the neutron-$\gamma$ discrimination of two recently developed plastic scintillators (including EJ299-33), of a recently introduced crystal (doped $p$-terphenyl), and of usual crystal and liquid scintillators.
We focus on the relation between the PSD properties of the scintillators and other characteristics such as pulse shapes and light output.
Using simple assumptions we obtain the quantitative influence of the signal shapes and total light on the figure of merit measuring the discrimination quality.
With these results, a detailed comparison of the scintillators can be performed and limitations of current organic scintillators for neutron-$\gamma$ discrimination can be discussed.

\section{Experimental details}
\subsection{Scintillators}

Cylindrical samples with a diameter of 5 cm and a height of 5 cm of the following scintillators were studied:

\begin{itemize}
\item EJ299-33 plastic (referred to as EJ299 in the following). The sample was wrapped with a reflective film.

\item A plastic prototype developed by the LIST/LCAE laboratory (CEA, France). Its composition is similar to that of sample 6 of Ref. \cite{Blanc}. The plastic was coated with EJ510 white reflective paint. It will be referred to as ``CEA-PS'' in the following.

\item A single crystal of the recently introduced doped $p$-terphenyl scintillator \cite{Budakovsky}. It was manufactured by the melt-grown technique and encapsulated by Cryos-Beta \cite{CryosBeta} in an aluminium container with white reflective paint applied on the inner face and a glass window for coupling to the photomultiplier tube (PMT).

\item A melt-grown trans-stilbene (stilbene) single crystal, also manufactured by Cryos-Beta. Its encapsulation is similar to that of the $p$-terphenyl crystal.

\item Cells of the NE213 and BC501A liquid scintillators, known to offer excellent PSD performance \cite{Moszynski}. The two liquids are equivalent in terms of discrimination performance and light yield \cite{Moszynski}. Including both of them allowed to check the stability and consistency of our results.

The NE213 sample was contained in a glass cell with windows on both ends for coupling to PMTs.  The outer face of the cell was coated with white reflective paint. During the measurements, a PMT was coupled to one window, while the other window was covered by a polytetrafluoroethylene (PTFE) reflector disk.

The aluminium cell containing BC501A was manufactured and filled by Saint-Gobain Crystals \cite{SaintGobain} (model 2MAB-2F2BC501A). This cell has two glass windows for coupling to PMTs, and the inner face of its wall is coated with white reflective paint. As for the NE213 cell, the unused window was covered with a PTFE disk.

\item A cell of the BC537 deuterated liquid scintillator, equivalent to NE230 and EJ315.
 The cell, manufactured by Saint-Gobain Crystals, is similar to the BC501A cell (model 2MAB-2F2BC537).
\end{itemize}

In order to eliminate possible variations due to physically different photomultiplier tubes, all scintillators were coupled alternatively to the same Hamamatsu R329-02 PMT equipped with the same voltage divider (assembly H7195).
The optical coupling between the scintillators and the PMT window was insured by BC630 optical grease.
The PMT was biased at the same high voltage for all scintillators and all measurements ($-1650$ V), to maintain the same response, in particular the same gain.

\subsection{Digital electronics}

The PMT anode signals were sent to a digital board developed at LPC Caen as part of the FASTER project \cite{FASTER}. The board is based on a digitizer with a 500-MHz sampling rate, an input range of $\pm1.2$ V and a 12-bit resolution. The bandwidth of 100 MHz is fixed by an anti-aliasing low-pass filter. An on-board FPGA can perform digital signal processing. In the present work it was used to perform baseline restoration (BLR).
The baseline-corrected digitized traces were written to disk for off-line analysis.

\subsection{Energy calibration}

The total light in a pulse was measured by the charge $Q$ obtained by integrating the whole digitized signal over a time gate beginning 20 ns before the maximum amplitude and with a duration of 600 ns. Varying the duration of this gate from 400 to 600 ns gave a charge stable to within 1 \%.

The total charge $Q$ was calibrated in terms of equivalent electron energy $E_e$ using the photoelectric peak of 59.5-keV $\gamma$-rays from $^{241}$Am and Compton electrons induced by $^{22}$Na and $^{137}$Cs $\gamma$-rays.
GEANT4 simulations were performed to determine what charge corresponded to the Compton edge energy. The simulated deposited energy distribution, including effects of multiple Compton scattering due to the scintillator finite size, was folded by a gaussian distribution representing the resolution, with a width adjusted to reproduce the experimental spectrum in the region of the Compton edge. The charge to be associated to the Compton edge energy corresponded to a fraction of 75 to 80 \% of the maximum height of the Compton distribution, depending on the scintillator \cite{TheseSenoville}.
A linear relation $Q = a (E_e - b)$ was assumed between the energy and the charge. The small energy-intercept $b$ is due to the non-linear response of organic scintillators below some 100 keVee \cite{Flynn,Horrocks}. It varied from 9 to 14 keVee depending on the scintillator, in agreement with previously reported values for organic scintillators \cite{Dietze}.

\subsection{Relative light yields}

The light yield $Y$ of a scintillator relative to that of BC501A, $Y/Y_\text{BC501A}$, was determined by the ratio of the slope $a$ of its calibration function to that of the BC501A cell.
The measured relative light yields $Y/Y_\text{BC501A}$ are given in Table \ref{TabYields}, together with light yields in photons/MeVee from the manufacturers and wavelengths of maximum emission $\lambda_{max}$. Relative uncertainties on our measured light yields were determined to be 5 \%.

The aim of this work is not to measure the absolute light yields of the scintillators.
Relative light yields are reported here merely for discussion of the scintillator properties in relation to neutron-$\gamma$ discrimination.
Although we used scintillators of the same shape and dimensions coupled to the same PMT and performed measurements in identical conditions for all of them, we expect our results to be affected by slightly different light collection efficiencies, due to the use of different reflectors, and by the different matching of the PMT response to the emission spectra of the scintillators.

On Tab. \ref{TabYields}, we note the good agreement of the relative light yields of $p$-terphenyl, NE213 and BC537 with values expected from manufacturer data. 
We observe a deviation of the stilbene relative light yield from the expected value. Since the wavelengths of maximum emission of stilbene and BC501A are different ($\lambda_{max}=390$ and 425 nm, respectively), we attribute this deviation to wavelength-dependent light collection effects (\textit{e.g.} reflectance of the reflector), and to the higher PMT quantum efficiency for stilbene than for BC501A.
The EJ299 plastic also shows a light yield relative to BC501A larger than indicated by manufacturer data. Because of its glass window, the BC501A cell has one additional optical interface between the scintillator and the PMT, compared to the EJ299 plastic coupled directly to the PMT. This is likely to lead to a larger light collection efficiency for EJ299.
Our EJ299 results might also be affected by performance variability, in favour of our sample, since this plastic was still under development as indicated by the manufacturer. In addition, we note that no electron light yield measurement relative to a known scintillator was reported in the characterizations of EJ299-33 \cite{PozziEJ299,NyibuleEJ299,CesterEJ299,LawrenceEJ299,WoolfEJ299}.

\section{Pulse shape discrimination}

\subsection{PSD procedure}

Data were taken with an AmBe source, without neutron moderator or $\gamma$-ray shield.
For neutron-$\gamma$ discrimination, we used the charge comparison method \cite{Heltsley}, in which the signal is integrated over two gates, one covering the full duration of the signal to give the total charge $Q$, and the other one delayed to integrate the pulse tail and thus give the slow charge $Q_s$.
The slow-to-total charge ratio $D=\frac{Q_s}{Q}$ was chosen here as the discriminating variable.

\subsection{Discrimination two-dimensional matrices}

\begin{figure}[htbp]
\begin{center}
\includegraphics[width=7cm]{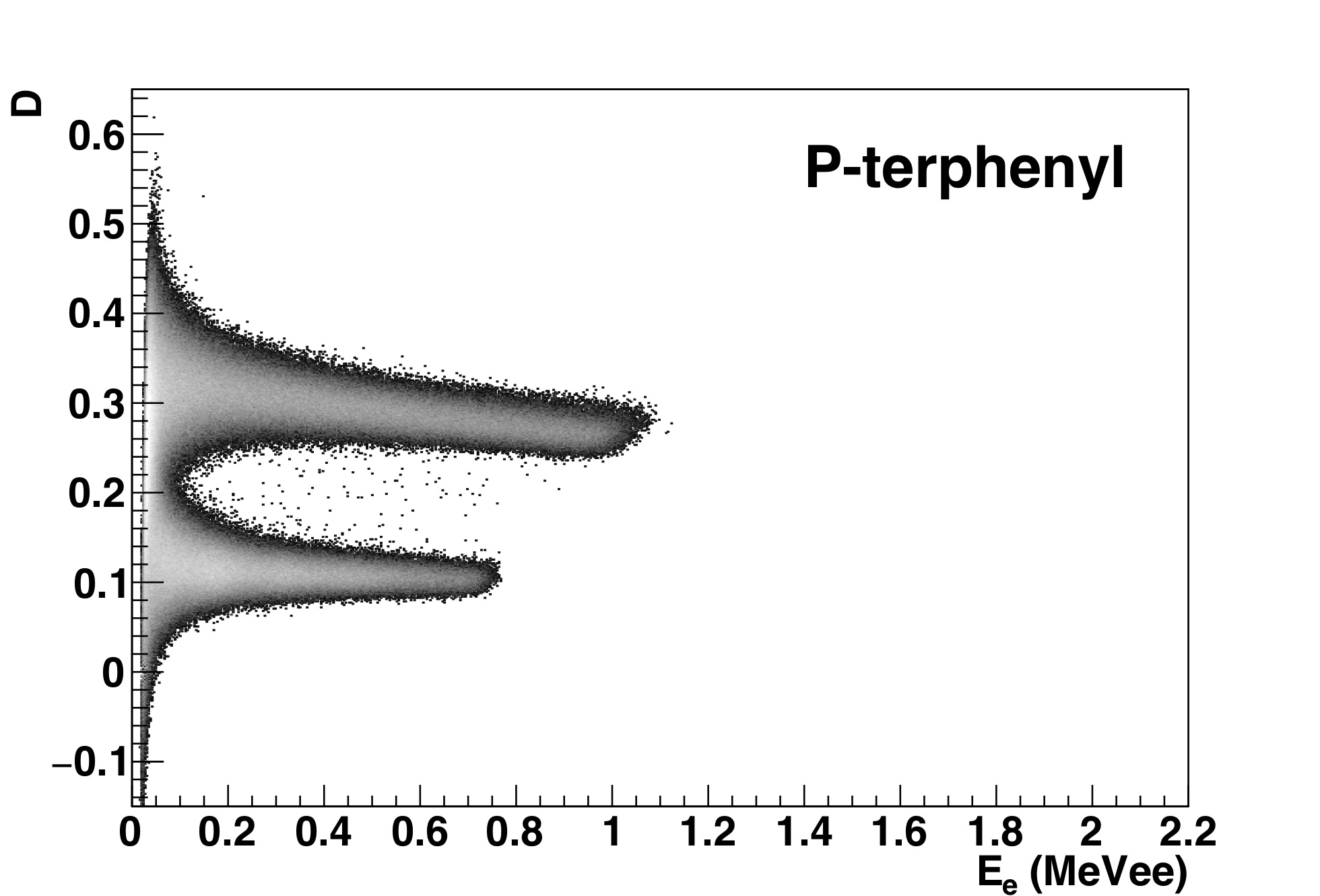}
\includegraphics[width=7cm]{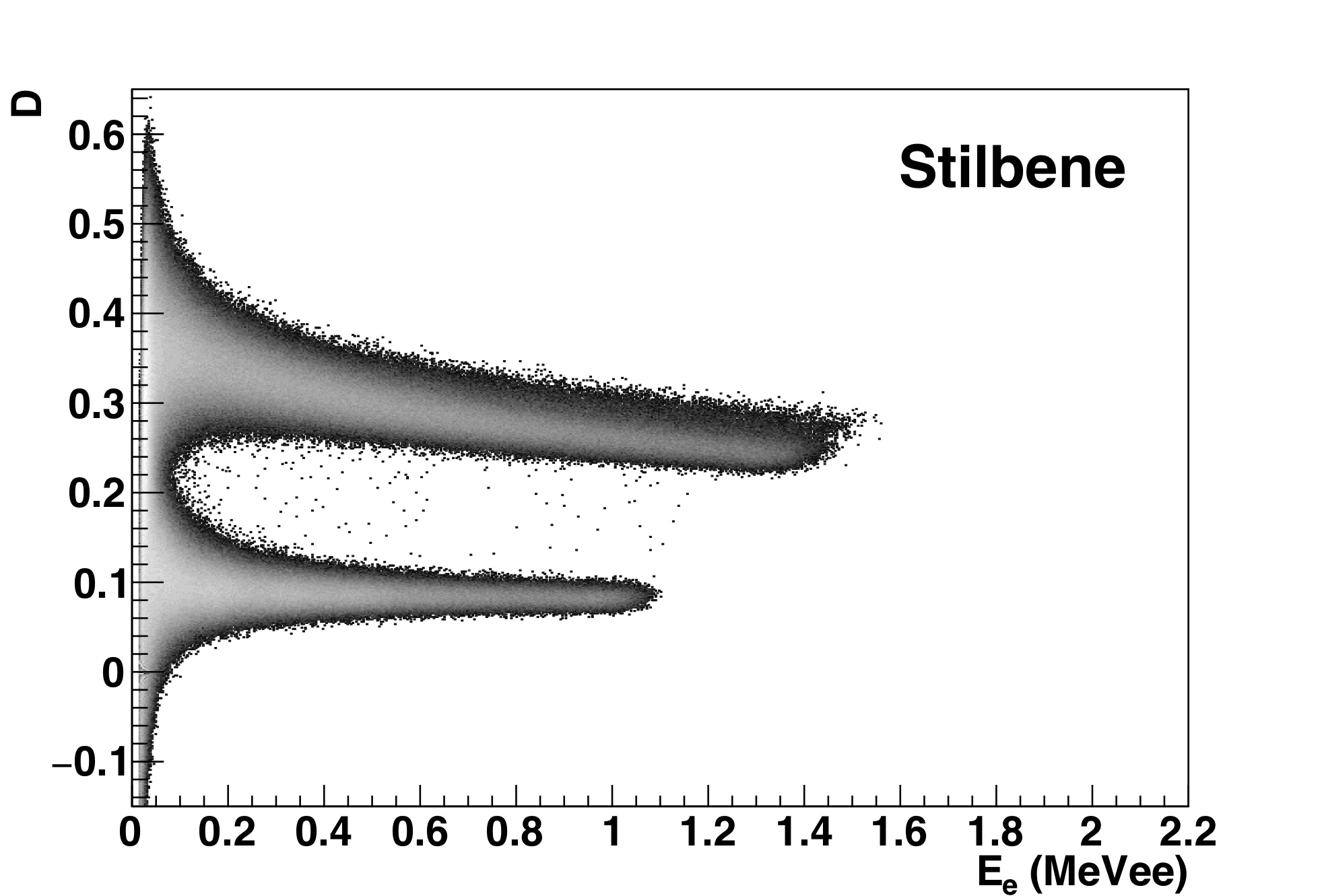}
\includegraphics[width=7cm]{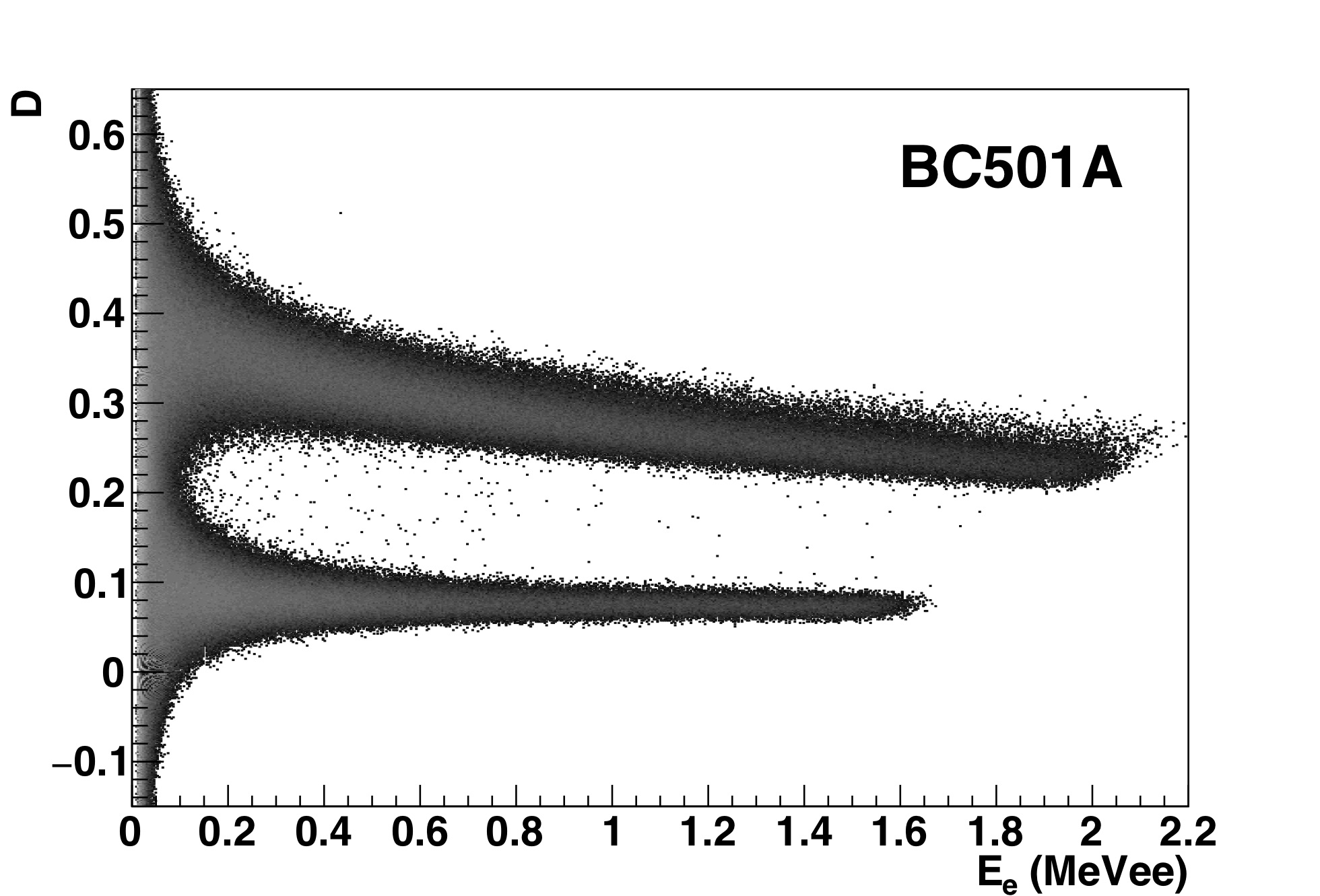}
\includegraphics[width=7cm]{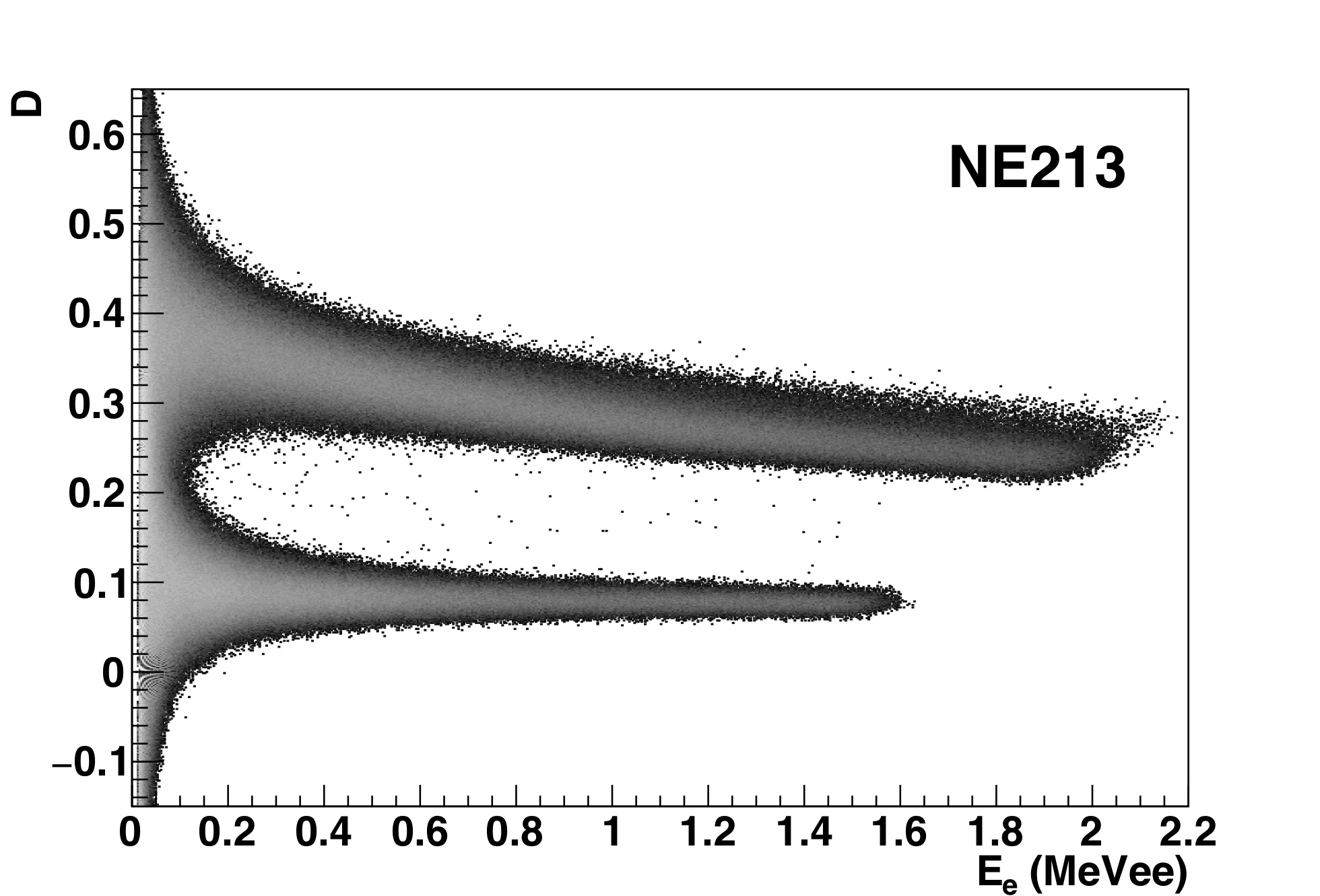}
\includegraphics[width=7cm]{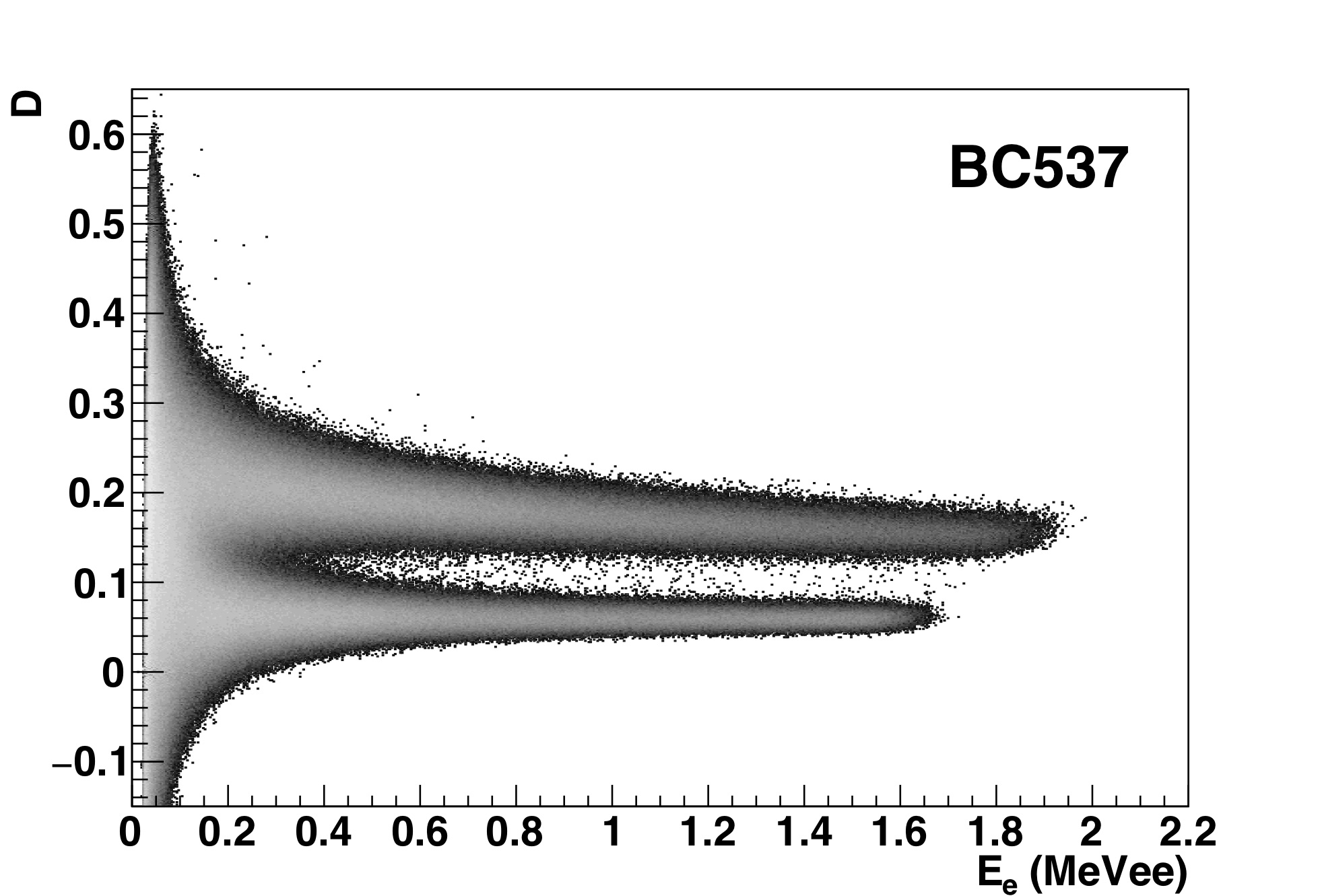}
\includegraphics[width=7cm]{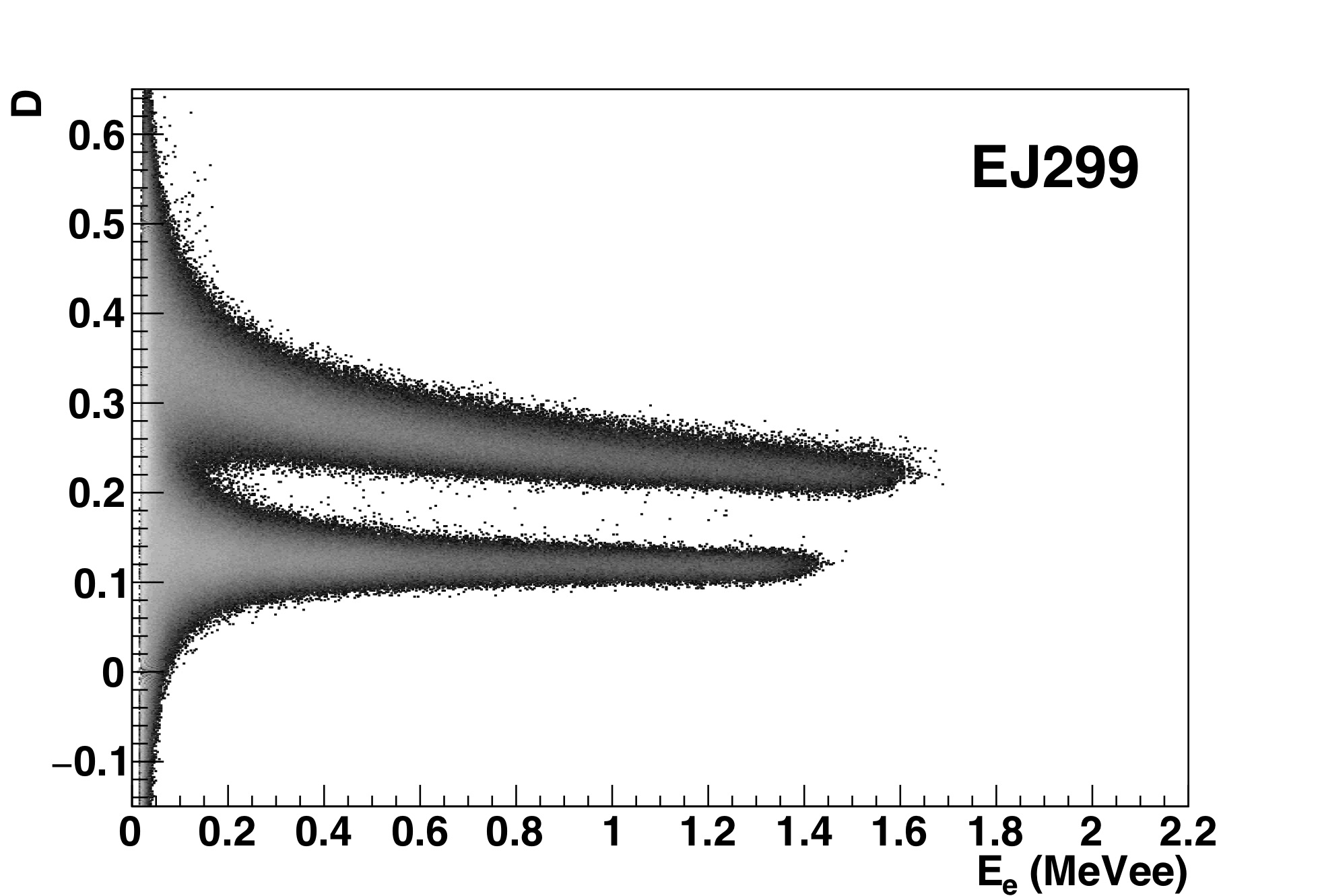}
\hspace*{-0.043\textwidth}
\includegraphics[width=15.9cm]{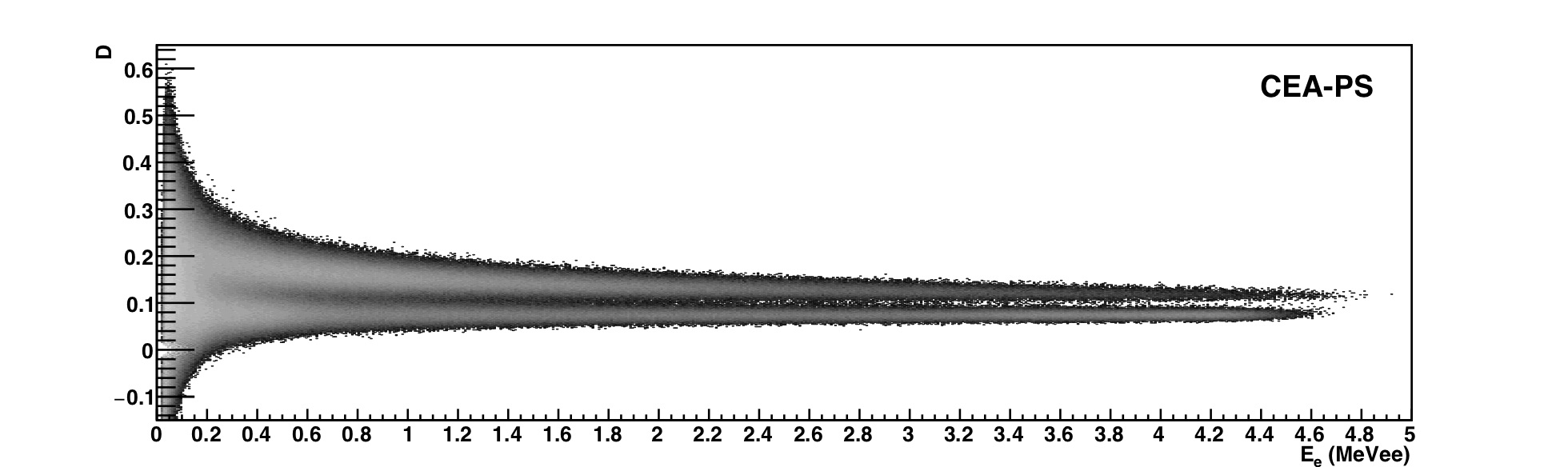}
\caption{Discrimination matrices showing the slow-to-total ratio $D$ as a function of the equivalent electron energy $E_e$, with a common $E_e$ scale. On each matrix, the branch with larger $D$ values corresponds to neutrons, while the other one corresponds to $\gamma$-rays.}
\label{FigHDisc}
\end{center}
\end{figure}

The two-dimensional identification matrices obtained by plotting the slow-to-total ratio $D$ as a function of the total charge $Q$ are shown on Fig. \ref{FigHDisc} for all the scintillators. They correspond to the best discriminations, obtained with gates optimized for each scintillator.
On each matrix, two branches are clearly visible, one at larger values of the slow-to-total ratio $D$ corresponding to neutrons, and the other one at lower $D$ values corresponding to $\gamma$-rays.
The endpoints of the two branches correspond to signals with a pulse height equal to the digitizer saturation voltage.
At a given equivalent electron energy $E_e$, \textit{i.e.} a given total charge, neutron signals have a smaller pulse height than $\gamma$-ray signals.
Therefore neutron pulses saturate the digitizer at a larger energy $E_e$ than $\gamma$-ray signals.
The differences in the endpoint energies from one scintillator to the other are due to different light output responses to electrons and neutron-induced recoils, mainly protons in the present neutron energy range (deuterons for BC537).

The identification matrices of Fig. \ref{FigHDisc} allow a first, qualitative, comparison between the discrimination capabilities of the various scintillators. The separation performed by the $p$-terphenyl and stilbene crystals and by the NE213-BC501A liquid scintillator appears to be better than that of the EJ299 and CEA-PS plastic scintillators and of the deuterated liquid BC537. The EJ299 plastic gives a more efficient discrimination than BC537, evidenced by the larger separation between the neutron and $\gamma$-ray branches, and by the lower energy at which the two branches start to overlap. Although its performance is limited, the CEA-PS plastic scintillator discriminates neutrons and $\gamma$-rays.

One notes that the slow-to-total ratio $D$ of the neutron branch decreases as the equivalent electron energy increases, while the $\gamma$-ray branch is characterized by a much more constant $D$ ratio.
These behaviours are visible for all scintillators, although less clearly for BC537 and CEA-PS.
Such evolutions of PSD parameters with energy are typical (see \textit{e.g.} Refs. \cite{Zaitseva2018,WoolfEJ299,MoszynskiLargeVolume,CesterPSDDig}).
They reveal the different dependence of neutron and $\gamma$-ray pulse shapes on the energy of the recoil particle: the neutron-induced average signal evolves significantly with the energy \cite{CesterPSDDig,Guerrero}, whereas the $\gamma$-ray pulse shape is very stable.

\subsection{Measurement of the discriminating quality}

\begin{figure}[htbp]
\begin{center}
\includegraphics[width=7cm]{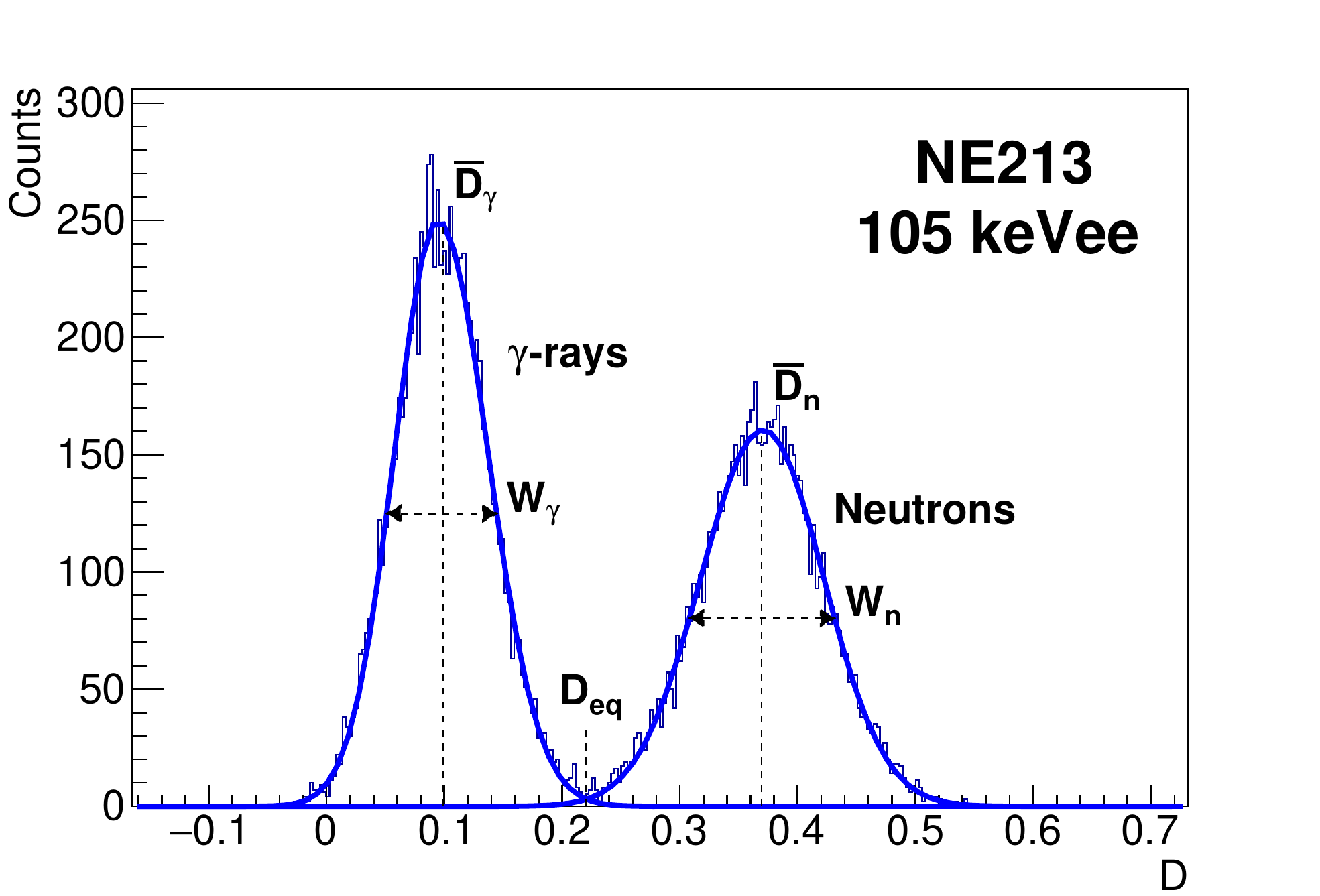}
\caption{Distribution of the discriminating variable $D$ obtained for NE213 in an energy interval of width $\Delta E_e=$ 3.8 keVee at $E_e=105$ keVee. The corresponding figure of merit is $M=1.25$. The full lines represent the two asymetrical gaussian functions resulting from the fit of the distribution.}
\label{FigProjY}
\end{center}
\end{figure}

To provide a quantitative comparison of the scintillators, the discrimination quality was measured as a function of the energy $E_e$ from the distributions of the discriminating variable $D$ in narrow energy intervals of width $\Delta E_e$ ($\Delta E_e /E_e$ from 3 to 6 \%).
Each interval contained at least 7500 events.
Intervals were taken over the whole energy range covered.
An example of such a distribution is shown on Fig. \ref{FigProjY} for NE213 in an energy interval of width $\Delta E_e=3.8$ keVee at $E_e=105$ keVee.
The discrimination quality at a given energy $E_e$ was measured by the figure of merit $M$ defined by \cite{BrooksReviewNIM}:
\begin{equation} \label{EqFoM}
M = \frac{\overline{D}_n-\overline{D}_\gamma}{W_n+W_\gamma},
\end{equation}
where $\overline{D}_n$ and $\overline{D}_\gamma$ are the mean positions of the neutron and $\gamma$-ray peaks, respectively, and $W_n$ and $W_\gamma$ are the corresponding Full Widths at Half Maximum (FWHM) of the peaks (see Fig. \ref{FigProjY}).
These means and widths were obtained from a fit of the distribution with a sum of two asymmetric gaussian functions, which gave a better description of the peaks than gaussian functions.
Each function was defined by:
\begin{equation} \label{EqAGauss}
f_i(D) = \left\{
\begin{array}{rl}
a_i \, \exp \! \left[-\frac{(D-D_{0i})^2}{2 {\sigma_{iL}}^2}\right] & \mbox{if } D < D_{0i}, \vspace{1mm} \\
a_i \, \exp \! \left[-\frac{(D-D_{0i})^2}{2 {\sigma_{iR}}^2}\right] & \mbox{if } D \ge D_{0i},
\end{array}
\right.
\end{equation}
where $i = n$ or $\gamma$, $a_i$ is the amplitude, and $D_{0i}$ is the most probable value, which is different from the mean value $\overline D_i$ since the function is asymmetric.
Defining $w_{iL}$ and $w_{iR}$ as the half FWHM, respectively in the domains $D < D_{0i}$ and $D > D_{0i}$, then $\sigma_{iL}$ and $\sigma_{iR}$ are given by $\sigma_{iL,R} = w_{iL,R}/1.177$.  The full width at half maximum is thus $W_{i} = \frac{2.354}{2} (\sigma_{iL} + \sigma_{iR})$.
The mean value is given by $\overline{D}_i = D_{0i} + \sqrt{\frac{2}{\pi}}\left( \sigma_{iR} -\sigma_{iL} \right)$ \cite{AsymGaussians}.

The figure of merit chosen here offers the advantage of depending only on the shape of the distribution of the discriminating variable.
It is appealing to characterize the discrimination performance with the contamination fraction of the distribution of one particle type by particles of the other type, but this measure depends on the experimental conditions, such as the $\gamma$-ray to neutron emission ratio of the source, the ratio of detection efficiencies for neutron and $\gamma$-rays, or the use of neutron moderators or $\gamma$-ray shields.

The total and slow integration gates were adjusted for each scintillator in order to maximise the figure of merit $M$.
As a general feature, we found that $M$ increased with the duration of the gates.
In practice, the integration of the signal was limited to 600 ns after the leading edge, which was the minimal duration of BLR operation.
Therefore the width of the total gate, used also for the energy calibration, was set to 600 ns.
The total gate began 20 ns before the maximum sample \textit{i.e.} a few ns before the leading edge.
The optimal start time of the slow gate varied from 12 to 54 ns after the maximum sample, depending on the scintillator, as presented on Table \ref{TabGatesEM1}.
The slow gate was set to end at the same time as the total gate.

\subsection{Comparison of discrimination figures of merit}

\begin{figure}[htbp]
\begin{center}
\includegraphics[width=8.8cm]{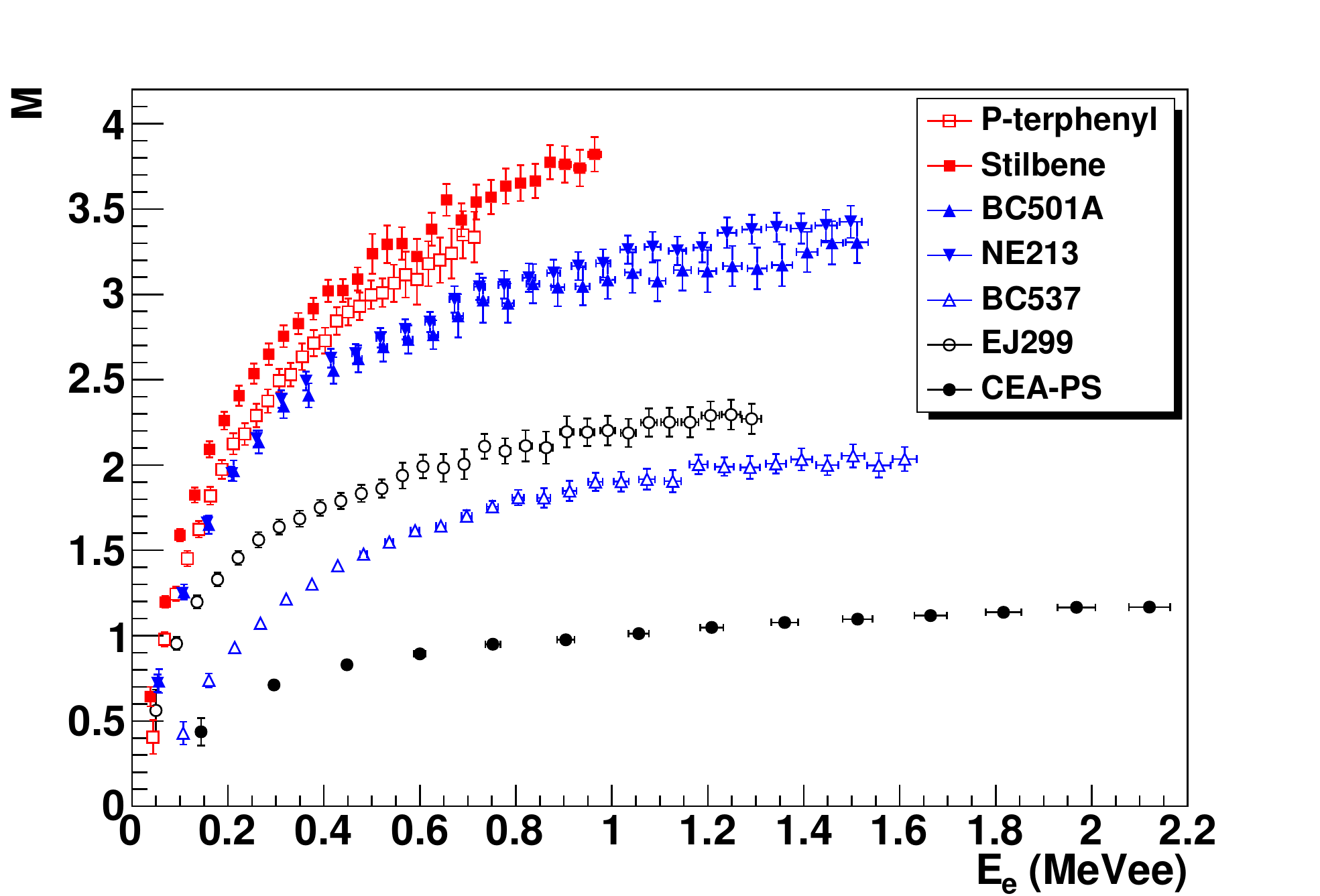}
\caption{Figures of merit $M$ as a function of the equivalent electron energy $E_e$. The error bars on $M$ are statistical, whereas those on $E_e$ show the widths of the intervals chosen to determine the values of $M$.}
\label{FigMQcal}
\end{center}
\end{figure}

Fig. \ref{FigMQcal} shows the figure of merit $M$ as a function of the equivalent electron energy $E_e$, for the different scintillators. The endpoint of each curve corresponds to the energy up to which the discrimination quality could be quantified, \textit{i.e.} to the endpoint energy of the $\gamma$-ray branch due to the digitizer saturation.

The largest figures of merit are obtained with stilbene, consistently with previous works reporting an excellent discrimination for this crystal \cite{Brooks1959}.
The doped $p$-terphenyl crystal also gives an excellent discrimination, with figures of merit lower by about 0.2-0.3 compared to stilbene.
From the fits of the neutron and $\gamma$-ray $D$ distributions, we estimate that this difference corresponds to 5 to 10 times fewer misidentified events for stilbene than for doped $p$-terphenyl at a given equivalent electron energy.

The cells of the NE213 and BC501A liquids give very similar figures of merit, as expected for these two equivalent scintillators.
The NE213 cell tends to give slightly larger $M$ values, which might be due to the different placement of the reflective paint for the two cells. Overall, the two liquids offer an excellent discrimination performance, although significantly lower than that of the two crystals.

While the figure of merit of doped $p$-terphenyl is close to that of stilbene at 700 keVee, it decreases more rapidly when the energy is reduced, and it is closer to the figure of merit of NE213-BC501A below 400 keVee.
Although the figure of merit of the EJ299 plastic is much lower than those of the crystals and of the NE213-BC501A liquid, it is larger than 1 for most of the energy range, which already corresponds to a very good discrimination. As such, this plastic offers a very interesting alternative to crystals and liquids.
The differences between the $M$ curves of the two crystals, the NE213-BC501A liquid and the EJ299 plastic become smaller as the energy decreases, suggesting a common limitation at lower energies.
Interestingly, although the discrimination with the BC537 deuterated liquid is good at energies above a few hundred keVee, it worsens more rapidly towards low energies than for the other scintillators.
The discriminating quality of the CEA-PS scintillator is much lower than that of the other scintillators, even if with $M\approx 1$ over a large fraction of the energy range its discrimination is already reasonable.

\subsection{Threshold energies for efficient discrimination}

When applying PSD on an event-to-event basis, a simple choice for the limit in slow-to-total ratio $D$ between the $\gamma$-ray and neutron peaks at a given energy $E_e$ is the value $D=D_{eq}$ at which the two distributions are equal, \textit{i.e.} $f_\gamma(D_{eq}) = f_n(D_{eq})$, as illustrated on Fig. \ref{FigProjY}.
With this choice, a given event is considered as a neutron if $D>D_{eq}$ and as a $\gamma$-ray if $D<D_{eq}$.
The value of $D_{eq}$ can be determined using the results of the fit of the $D$ distribution.
Then, it is interesting to evaluate the number of misidentified events.
Integrating the neutron distribution function $f_n$ for $D < D_{eq}$ gives the number of misidentified neutrons $N_{n \rightarrow \gamma}$, and integrating the $\gamma$-ray distribution function $f_\gamma$ for $D > D_{eq}$ gives the number of misidentified $\gamma$-rays $N_{\gamma \rightarrow n}$.
The numbers of neutrons and $\gamma$-rays, respectively $N_n$ and $N_\gamma$, can be obtained by integrating each of the two distribution functions over the full $D$ range.
We determined that, in our conditions, a figure of merit $M=1$ corresponds to fractions of misidentified neutrons $N_{n \rightarrow \gamma}/N_n$ and of misidentified $\gamma$-rays $N_{\gamma \rightarrow n}/N_\gamma$ of the order of 1 \%, for all scintillators. This is a useful reference point that might be considered as a threshold for ``good'' discrimination, even if in practice a particular experiment or application might accept lower or higher fractions of misidentified events. This fraction of misidentified events should not be taken as universal, as
it depends on experimental details affecting the ratio of detected neutrons and $\gamma$-rays.
Below and above $M=1$, the fractions of misidentified events evolve very rapidly with $M$, increasing (decreasing) by roughly a factor 10 for a decrease (increase) of $M$ of 0.4.

The ``threshold'' equivalent electron energies at which $M=1$ are given in Tab. \ref{TabGatesEM1} for the different scintillators, together with the corresponding neutron minimum energies computed with the proton (deuteron for BC537) light output functions indicated in the table.
Such a function gives the energy of an electron that produces the same total light as the considered proton or deuteron.
As mentioned above, the response of an organic crystal to a heavy particle depends on the direction of the particle with respect to the crystallographic axes \cite{Brooks1974}.
For our $p$-terphenyl and stilbene crystals the cylinder axis, along which the neutrons were incident, corresponded to the direction of minimum light output response (artificial $c^\prime$ axis perpendicular to the cleavage plane \cite{Brooks1974}).
We used light output functions in agreement with these conditions of neutron incidence \cite{HansenRichterStilbene, SardetPTerphenyl}.
Furthermore, the $p$-terphenyl light output function that we used was determined with a sample from the same manufacturer \cite{SardetPTerphenyl}.
The light output function of the CEA-PS scintillator was assumed to be similar to that of a typical plastic, NE102 (equivalent to BC400 or EJ200).

Uncertainties on the ``threshold'' energies include effects of the uncertainties on the figure of merit $M$ and on the energy calibration parameters, as well as possible systematic effects of a non linear light output response to electrons below some 100 keVee \cite{Flynn,Horrocks}.
For CEA-PS, the uncertainty on the equivalent electron energy is larger and dominated by the error on $M$, due to its much weaker energy dependence.

We observe that the ``threshold'' electron energies are consistent with the hierarchy of figures of merit, with stilbene showing the lowest electron energy, followed by $p$-terphenyl.
However, the more efficient proton light response of $p$-terphenyl compensates this difference when the electron energy is translated into neutron threshold energy.
The neutron threshold energies of $p$-terphenyl and stilbene are 492(22) and 517(25) keV respectively, with a difference of 25 keV smaller than the combined error of 34 keV.
Therefore, although giving an overall better discrimination than $p$-terphenyl, stilbene does not present a significantly lower neutron threshold.

As noted above, neutrons were incident on the crystals along the direction of their minimum response.
We did not take data with incidence along the direction of maximum response, as on the one hand the precise determination of this direction requires a dedicated measurement, and on the other hand this would not favour one of them since the response anisotropy is very similar for stilbene and $p$-terphenyl \cite{Brooks1974}.

NE213 and BC501A give identical thresholds, in agreement with their very similar discrimination performance. The electron energy threshold of the two liquids is 40 \% larger than that of stilbene. Similarly to $p$-terphenyl, this difference is partially compensated by the more efficient light response of NE213-BC501A to protons. The neutron threshold energy obtained for the two liquids is around 550 keV.

Even if the threshold electron energy of EJ299 is only 20 \% larger than for NE213-BC501A, the neutron threshold is 900 keV, 60 \% larger than for the liquid, due to a less favourable response to protons.

As noted above, the figure of merit decreases faster at low energies for the deuterated liquid BC537 than for the other scintillators. The $M=1$ neutron energy of around 1550 keV for BC537 is 70 \% higher compared to EJ299 and a factor 3 larger compared to the crystals and the NE213-BC501A liquid.

The figure of merit of the CEA-PS scintillator reaches $M=1$ at 1 MeVee, corresponding to an estimated neutron energy threshold of 3 MeV.

\subsection{Effects of pulse shapes and total light on discrimination}

We consider the fluctuations responsible for the FWHM $W_n$ and $W_\gamma$ of the neutron and $\gamma$-ray peaks in the expression of the figure of merit $M$ (Eq. \ref{EqFoM}).
We assume on the one hand that $W_n$ and $W_\gamma$ are proportional to the respective standard deviations $\sigma_{D_n}$ and $\sigma_{D_\gamma}$ of the neutron and $\gamma$-ray distribution functions, and on the other hand that the total and slow charges $Q$ and $Q_s$ are proportional to the numbers of photoelectrons in the total and slow integration gates, respectively $N$ and $N_s$.
Since the measurements were made in narrow gates on $Q$, $N$ can be considered constant and $N_s$ is expected to follow a binomial distribution, with a standard deviation given by $\sigma_{N_s} = \sqrt{N\overline{D}(1-\overline{D})}$ \cite{MoszynskiLargeVolume}.

With the above assumptions, one obtains:
\begin{equation} \label{EqFoMDecomp}
M \propto \sqrt{Q} \,\, \frac{\overline{D}_n - \overline{D}_\gamma} {\sqrt{\overline{D}_n\left(1-\overline{D}_n\right)} + \sqrt{\overline{D}_\gamma\left(1-\overline{D}_\gamma\right)}}.
\end{equation}

We define the \textit{pulse shape figure of merit} as:
\begin{equation} \label{Eqm}
m = \frac{\overline{D}_n - \overline{D}_\gamma} {\sqrt{\overline{D}_n\left(1-\overline{D}_n\right)} + \sqrt{\overline{D}_\gamma\left(1-\overline{D}_\gamma\right)}}.
\end{equation}
This term gives the quantitative effect of the neutron and $\gamma$-ray pulse shapes, via their average slow-to-total ratios, on the figure of merit $M$.
Therefore the figure of merit $M$ can be decomposed as the product of the pulse shape dependent term $m$ and a pulse shape independent term proportional to $\sqrt{Q}$ quantifying the effects of the fluctuations of the total light $Q$. One expects the pulse shape figure of merit $m$ to show some variation with the total light $Q$ due to the dependence of the signal shapes on the energy.

\begin{figure}[htbp]
\begin{center}
\includegraphics[width=8.8cm]{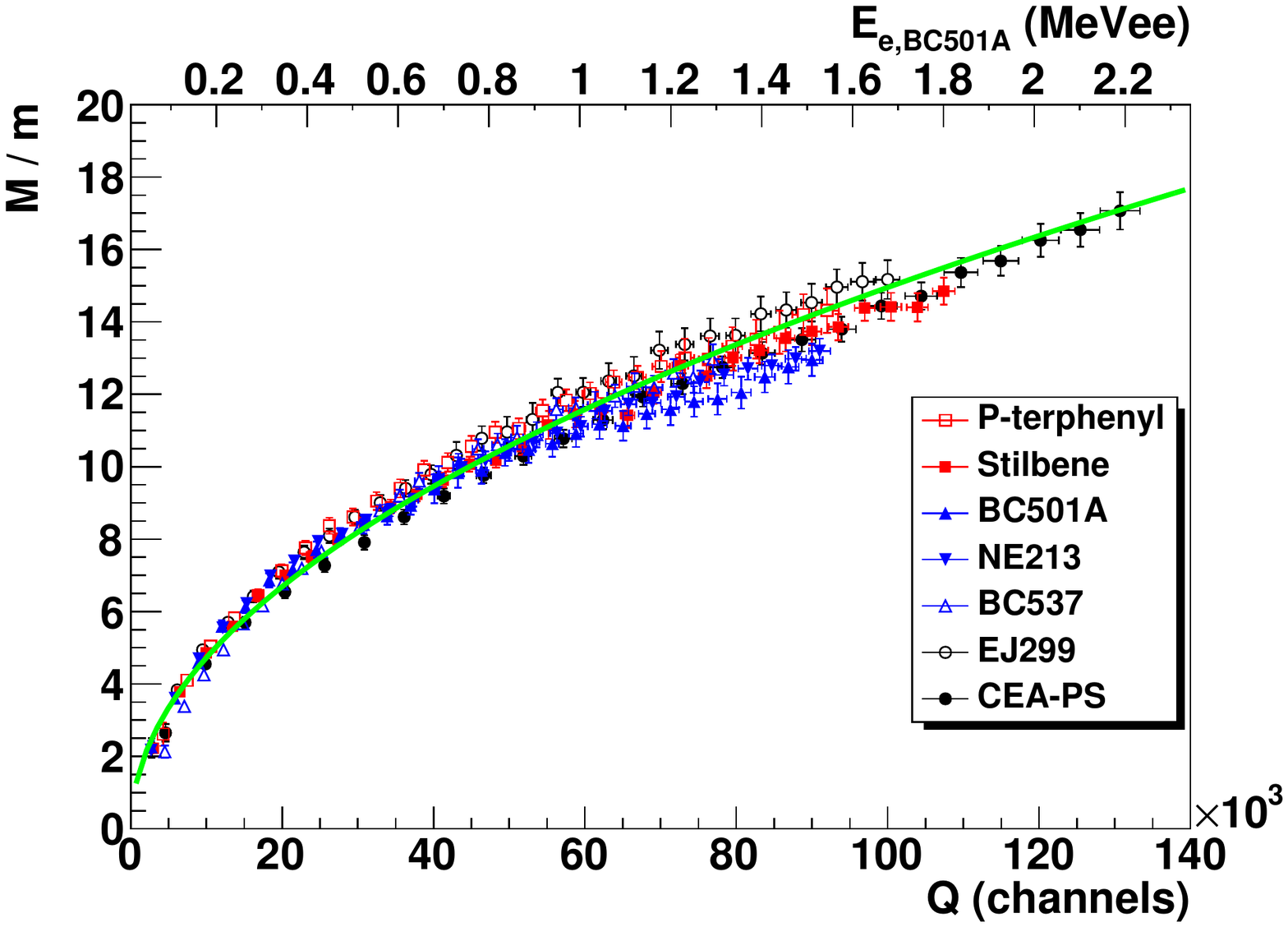}
\caption{Ratio $M/m$ of the figure of merit $M$ and pulse shape figure of merit $m$
as a function of the raw total charge $Q$. The green full line shows the $k\sqrt{Q}$ function with $k=4.73\times 10^{-2}$ a.u.. The upper horizontal axis gives the equivalent electron energy in BC501A.}
\label{FigRenMQraw}
\end{center}
\end{figure}

The ratio of the figure of merit $M$ and the pulse shape figure of merit $m$ is expected to be proportional to $\sqrt{Q}$ and, as such, should be independent of the scintillator.
Fig. \ref{FigRenMQraw} shows the $M/m$ ratio as a function of $Q$, with $M$ and $m$ obtained from Eq. (\ref{EqFoM}) and (\ref{Eqm}), respectively.
The charge $Q$ is used rather than the equivalent electron energy $E_e$ since the latter calibration is specific to each scintillator.
In particular, due to different light yields, a given energy $E_e$ in two scintillators does not correspond to the same amount of light.
For reference, the upper axis on Fig. \ref{FigRenMQraw} gives the equivalent electron energy in BC501A $E_{e,\text{BC501A}}$.
The full line shows the function $M/m = k\sqrt{Q}$, where the parameter $k$ was fitted to all data points, giving $k=4.73\times10^{-2}$ a.u..
The differences between the values of $M/m$ of the various scintillators are considerably reduced compared to the differences between the global figures of merit $M$, indicating a common behaviour for the evolution of $M/m$ as a function of the total light, as expected.
Furthermore, the experimental points are well described by the $M/m = k\sqrt{Q}$ relation, in agreement with expectations from the decomposition of $M$.
This shows that sources of fluctuations of $D$ other than the statistical fluctuations of the number of photoelectrons also give contributions proportional to $\sqrt{Q}$, and that the effects of such fluctuations on the discrimination with different scintillators are nearly identical.
For different scintillators, the $\sqrt{Q}$ term gives the effect of the light yields on $M$ at a given energy.
Therefore, differences of discrimination quality between the scintillators can be attributed only to differing pulse shape effects and light yields. This behaviour is common to all types of organic scintillators, crystals, plastics and liquids.

\begin{figure}[htbp]
\begin{center}
\includegraphics[width=8.8cm]{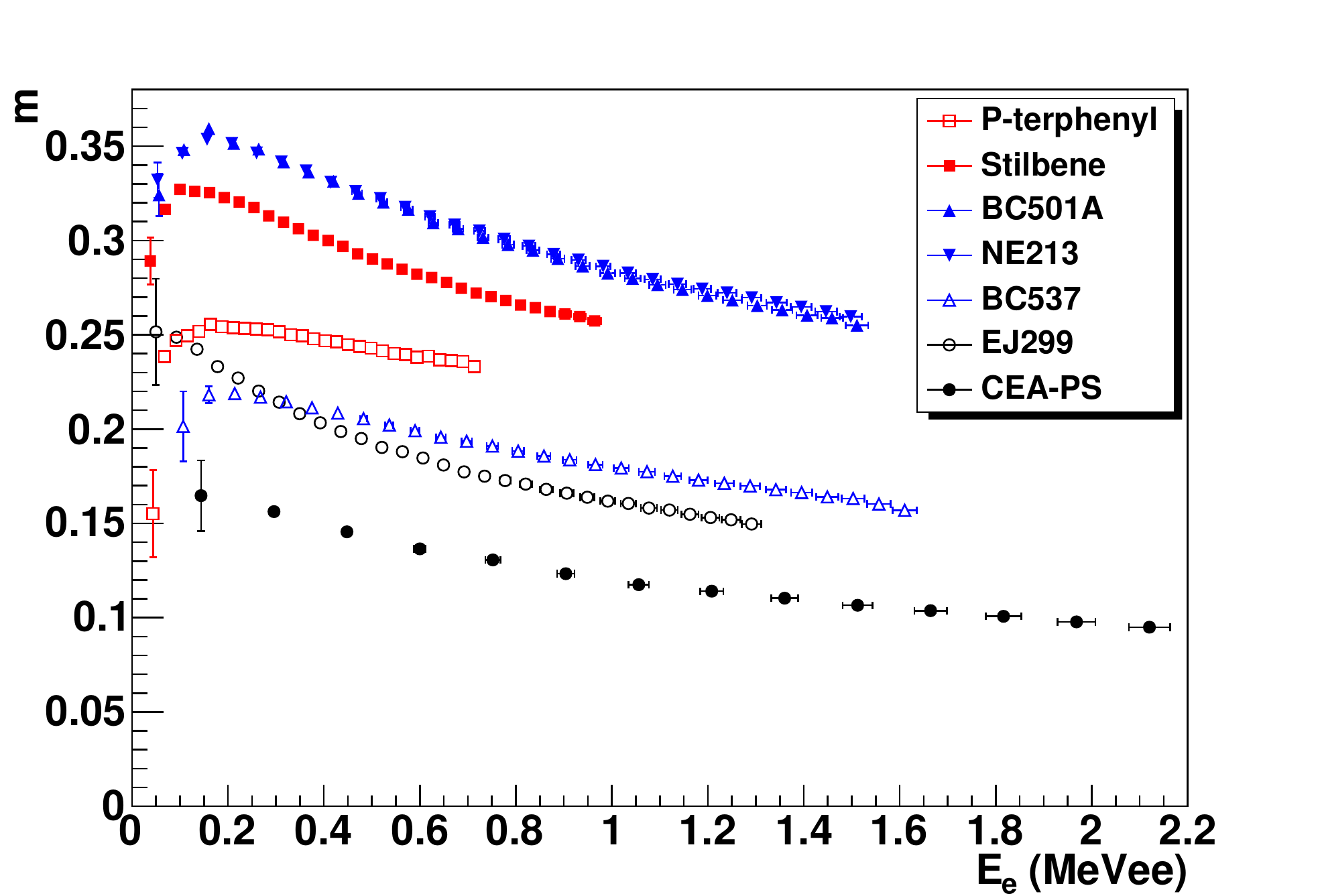}
\caption{Pulse shape figures of merit $m$ as a function of the equivalent electron energy $E_e$.}
\label{FigFomShapeQcal}
\end{center}
\end{figure}

The pulse shape figures of merit $m$ computed from the measured $\overline{D}_n$ and $\overline{D}_\gamma$ are presented on Fig. \ref{FigFomShapeQcal} as a function of the equivalent electron energy $E_e$, and in Tab. \ref{TabPSD300keV} for $E_e=300$ keVee.
Fig. \ref{FigFomShapeQcal} shows that BC501A and NE213 present the most favourable pulse shapes for discrimination, although their overall discrimination quality is lower than that of stilbene and $p$-terphenyl.
In addition, the two liquids show almost identical $m$ values throughout the energy range, as expected from their equivalence.
The pulse shape figures of merit of stilbene are only about 10 \% smaller than those of NE213-BC501A, and show an almost identical evolution with energy.
For doped $p$-terphenyl, $m$ takes much smaller values (20-30 \%) than for NE213-BC501A but decays at a slower rate with energy.
The other scintillators, BC537, EJ299, and CEA-PS, offer pulse shapes less favourable for discrimination than those of NE213-BC501A and the crystals.
Pulse shape figures of merit $m$ of BC537 and EJ299 have similar values, but they evolve differently with energy.
For EJ299, $m$ increases faster when the $E_e$ decreases, and becomes larger than for BC537 at 300 keVee and comparable to $m$ of $p$-terphenyl at 100 keVee.
This is consistent with the increase of the difference between the figures of merit $M$ of EJ299 and BC537 when the energy becomes smaller.
For the EJ299 plastic, $m$ is larger than for the other plastic CEA-PS.

All curves show an overall increase of $m$ when the energy $E_e$ decreases, indicating that pulse shapes become more favourable for discrimination at lower energies. This evolution is dominated by the increase of the neutron slow-to-total ratio $\overline{D}_n$.

The $\sqrt{Q}$ term evolves considerably over the energy range, while $m$ shows only moderate variations in comparison. Therefore the evolution of the figure of merit $M$ with energy is dominated by the $\sqrt{Q}$ dependence.
This term is responsible for the overall reduction of the figure of merit $M$ when the energy decreases, whereas the saturation of $M$ at higher energies is the combined effect of the milder rise of $\sqrt{Q}$ and the slow decrease of $m$.

\begin{figure}[htbp]
\begin{center}
\includegraphics[width=8.8cm]{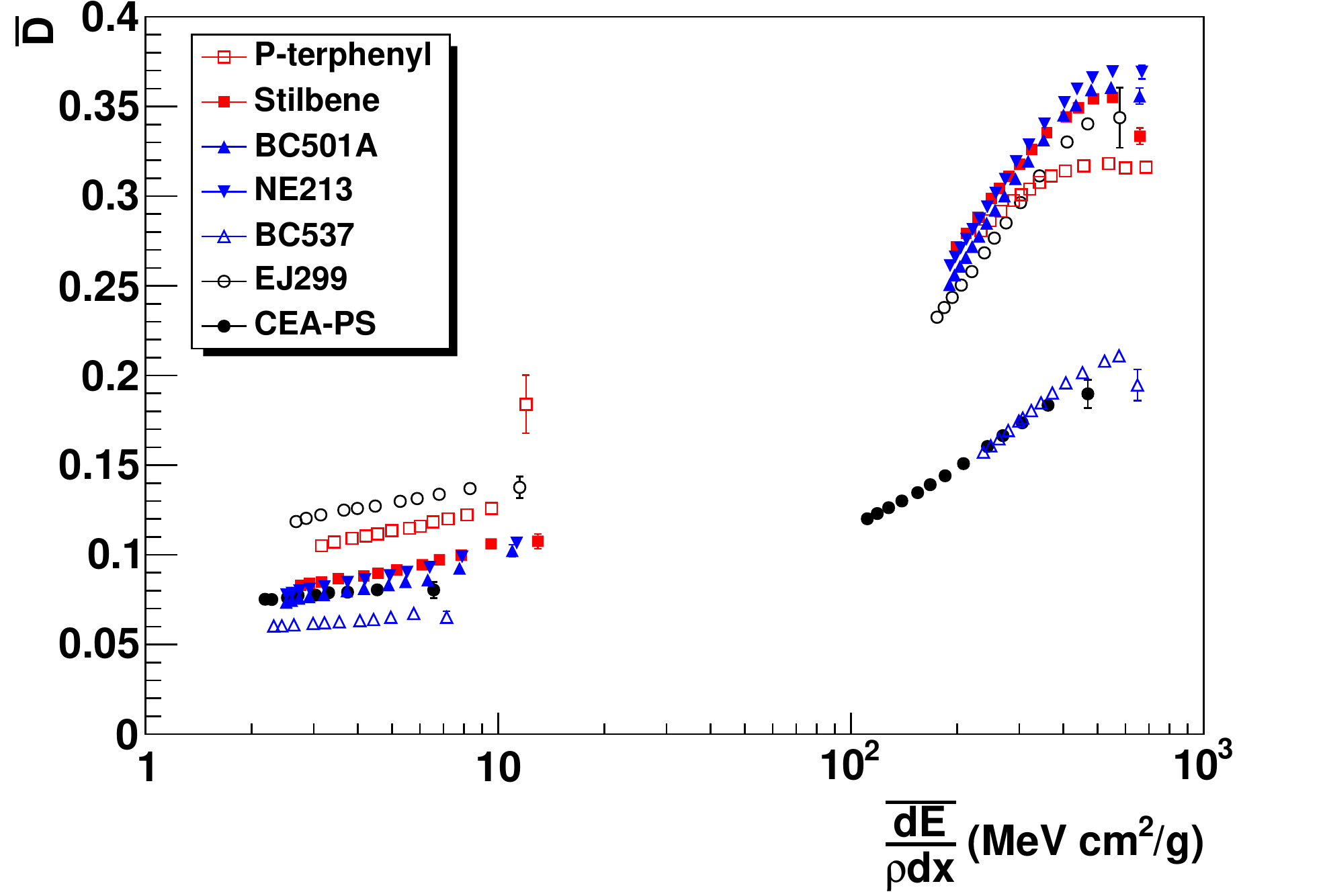}
\caption{Slow-to-total ratio as a function of the average mass stopping power of the recoil particle. Stopping powers smaller than 20 MeV cm$^2$/g correspond to electrons while those larger than 100 MeV cm$^2$/g correspond to recoil protons (deuterons in BC537).}
\label{FigDStopping}
\end{center}
\end{figure}

As mentioned above, the sensitivity of the scintillation pulse shape to the type of exciting particle is attributed to its dependence on stopping power.
To obtain the evolution of the average slow-to-total ratio $\overline{D}$ characterizing the pulse shape as a function of the stopping power, the latter was calculated from the measured equivalent electron energy $E_e$.
For data in the $\gamma$-ray branch, an electron with an energy $E_e$ was assumed. For the neutron branch, a recoil proton (deuteron for BC537) was considered, with an energy obtained from $E_e$ using the light output function of the scintillator.
To obtain a stopping power value representative of the scintillation produced over the recoil particle trajectory, the stopping power $dE/dx$ was weighted by the specific scintillation emission $dL/dx$ and averaged over the path of the particle:
\begin{equation}
\overline{\frac{dE}{dx}} = \frac{\int \frac{dL}{dx}\frac{dE}{dx}dx}{\int \frac{dL}{dx} dx} = \frac{\int \frac{dE/dx}{1+kB\,dE/dx} dE}{\int \frac{dE}{1+kB\,dE/dx}},
\end{equation}
where Birks' relation $\frac{dL}{dx} = \frac{S\,dE/dx}{1+kB\,dE/dx}$ was assumed between the specific scintillation emission and the stopping power. An ionization quenching parameter $kB = 9$ mg cm$^{-2}$ MeV$^{-1}$ typical of organic scintillators was used \cite{Birks64}.
Taking different values of this parameter for the various scintillators did not affect the results significantly.
Fig. \ref{FigDStopping} presents the measured slow-to-total ratio $\overline{D}$ as a function of the average mass stopping power of the recoiling particle $\overline{\frac{dE}{\rho dx}}$.
The points lower than 20 MeV cm$^2$/g correspond to the $\gamma$-ray branch, while those above 100 MeV cm$^2$/g correspond to the neutron branch.
The global trend of the curves is similar: while $\overline{D}$ is small and evolves slowly at low stopping powers typical of recoil electrons,
it shows a stronger variation and reaches large values for high stopping powers characteristic of neutron-induced recoils.
This dependence of the pulse shape on the stopping power of the recoil particle is the origin of the PSD capability as well as the different evolutions of the
neutron and $\gamma$-ray pulse shapes with energy revealed by PSD plots, \textit{e.g.} Fig. \ref{FigHDisc}.
We note that the weaker energy dependence of the neutron branch of BC537 and CEA-PS on Fig. \ref{FigHDisc} is associated to the smaller increase of their slow-to-total ratio $\overline{D}$ at high stopping powers.
All curves tend to saturate when the stopping power approaches 700 MeV cm$^2$/g, except $p$-terphenyl for which the saturation occurs at $\approx$ 400 MeV cm$^2$/g.
The values of $\overline{D}$ at low and high stopping powers appear specific to each scintillator.
This is attributed to the various compositions and structures of the scintillators, leading to a different interplay between photophysical properties and the prompt and delayed scintillation components.
Considering Fig. \ref{FigFomShapeQcal} and Fig. \ref{FigDStopping}, it is clear that the largest pulse shape figures of merit $m$ are obtained with scintillators showing the strongest evolution of $\overline{D}$ with the stopping power.

One interesting feature of Fig. \ref{FigDStopping} is the very similar evolution of $\overline{D}$ for stilbene and NE213-BC501A.
However, the pulse shape figure of merit $m$ of stilbene is systematically smaller. This appears to be due to the less efficient light output function of stilbene for protons: for a given energy $E_e$, the energy of a recoil proton in stilbene is larger than in NE213-BC501A, resulting in a smaller stopping power and a smaller $\overline{D}_n$, which in turn leads to a smaller $m$ (\textit{e.g.} see Table \ref{TabPSD300keV} for $E_e = 300$ keVee).

At a given equivalent electron energy $E_e$, a recoil deuteron in BC537 and a recoil proton in a hydrogenated scintillator have similar energies, therefore the stopping power of the deuteron in BC537 is larger. This is illustrated in Table \ref{TabPSD300keV} where the mass stopping power for a deuteron in BC537 is 25 to 50 \% larger than for a recoil proton in the other scintillators. In addition, the electron stopping power is smaller in BC537 than in the other scintillators, which is due to the smaller overall $Z/A$ ratio of BC537.
One would therefore expect $\gamma$-ray and neutron pulse shapes to differ more in BC537 and its discrimination capability to be better compared to hydrogenated scintillators. However, Fig. \ref{FigDStopping} shows that the dependence of the pulse shape on the stopping power is less pronounced in BC537 than in hydrogenated scintillators, which results in a smaller pulse shape figure of merit.
We also attribute this to the different photophysical properties of the scintillators. In this respect, we note that BC537 is based on deuterated benzene while the NE213-BC501A liquid is based on xylene. In particular, benzene is known to be a less efficient scintillation solvent \cite{Birks64}.

\subsection{Average signals}

\begin{figure}[htbp]
\begin{center}
\includegraphics[width=7cm]{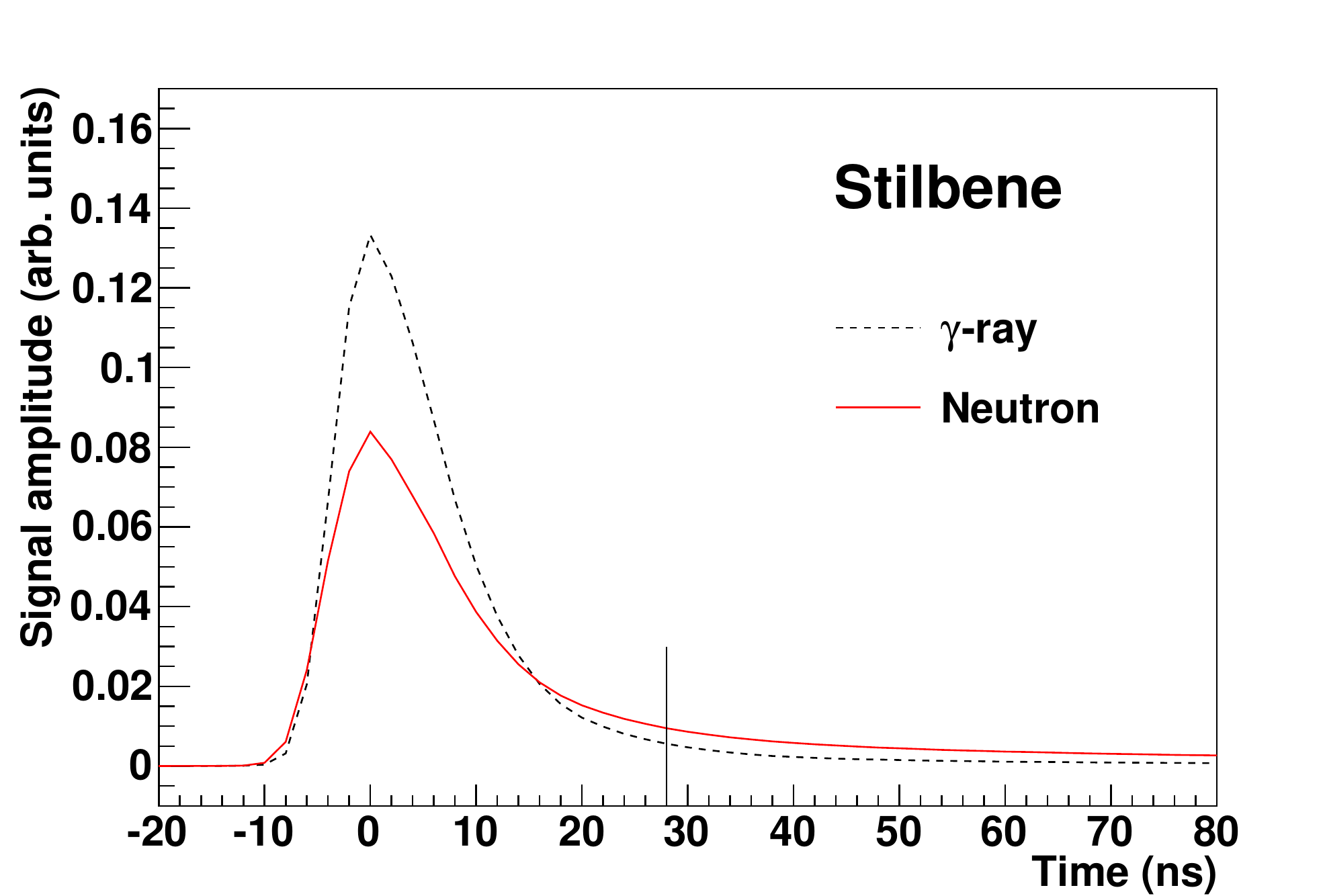}
\includegraphics[width=7cm]{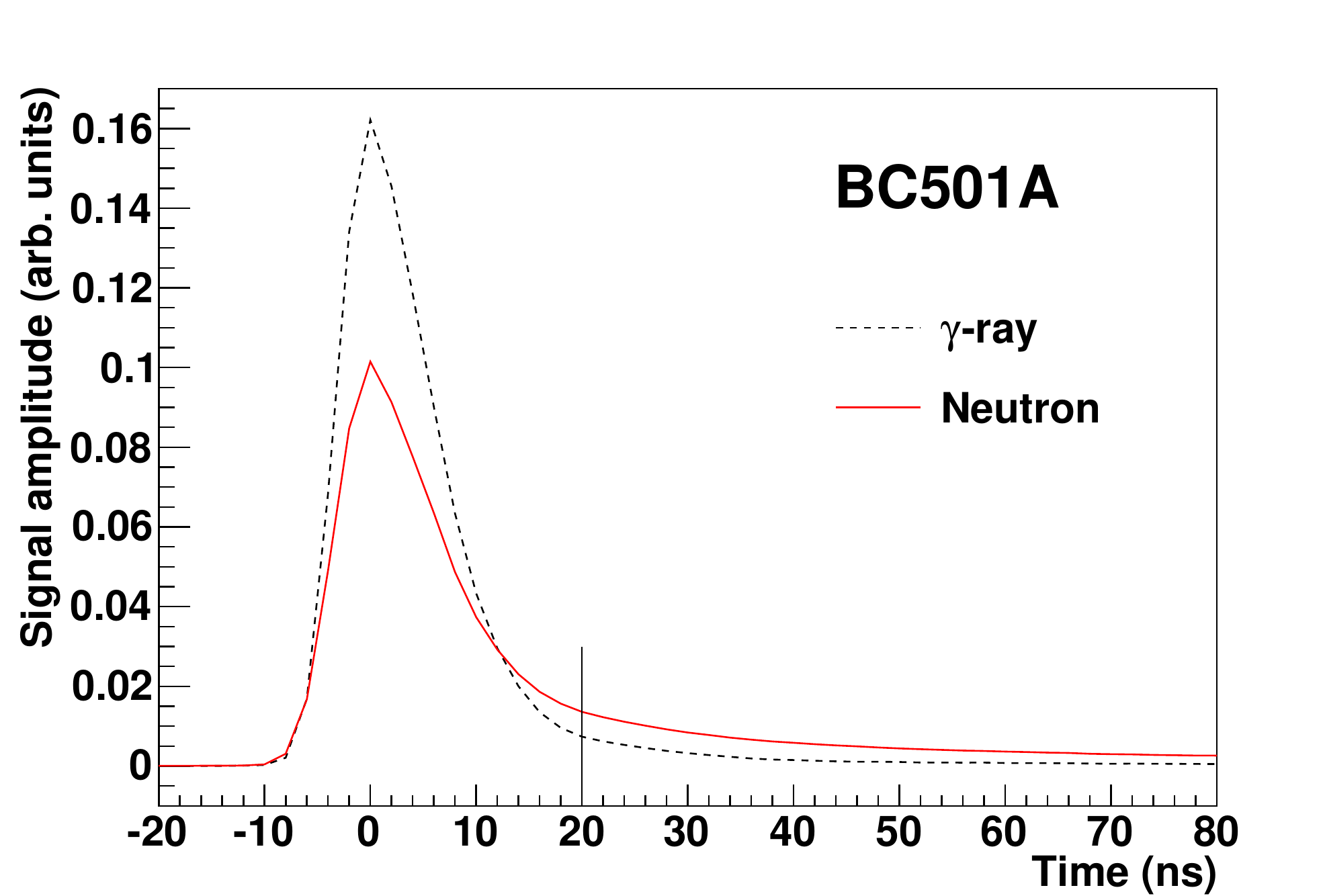}
\includegraphics[width=7cm]{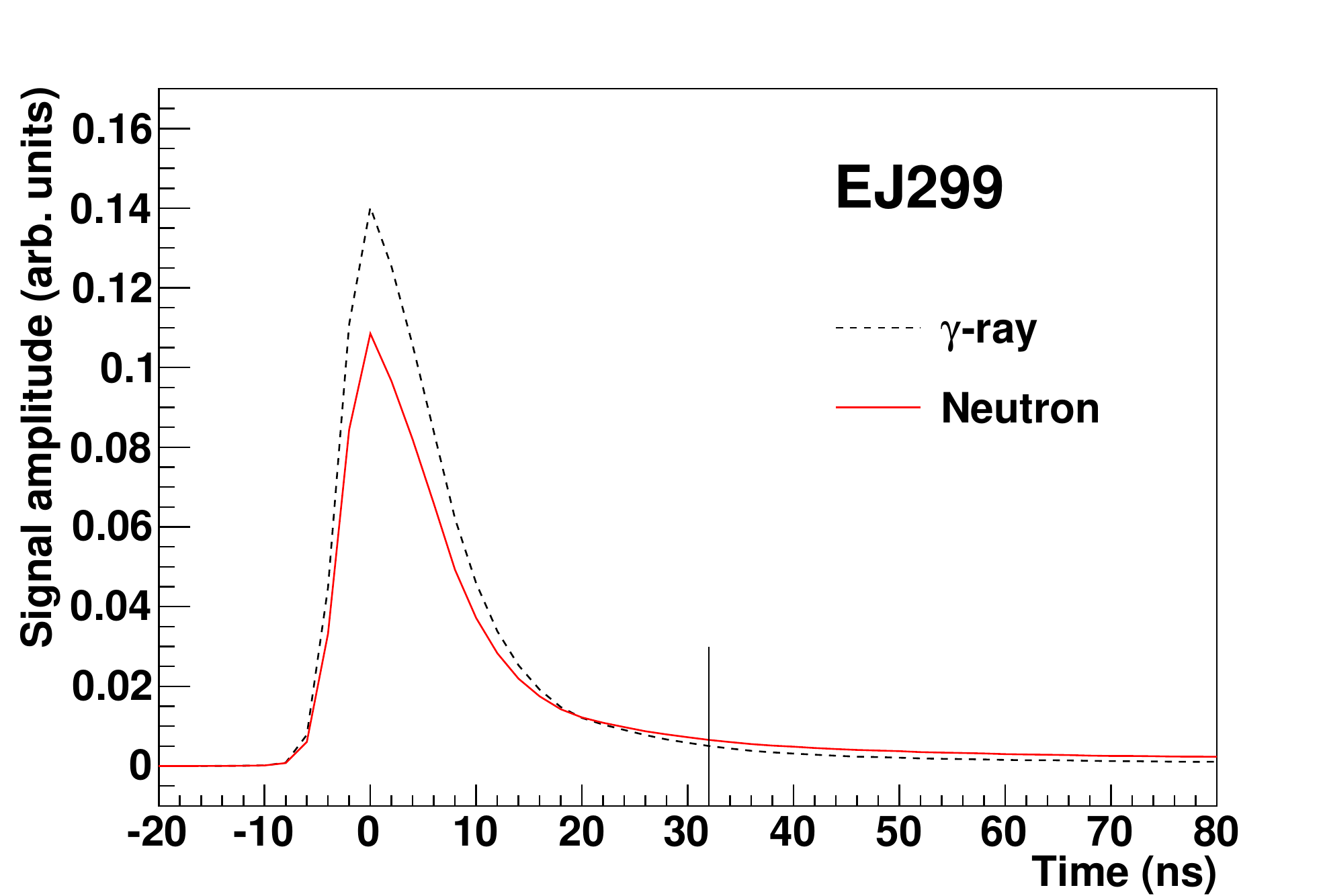}
\caption{Average signals of stilbene, BC501A, and EJ299 in the energy range 300-310 keVee, with a total integral normalized to 1, and shown in the [-20, 80] ns time range. The time of the maximum sample was taken as $t=0$. Statistical errors (not shown) are of the order of the thickness of the lines. The vertical full line shows the optimised start time of the slow integration gate, obtained by maximizing the figure of merit $M$.}
\label{FigAvSignals80nsBC501AStil}
\end{center}
\end{figure}

The digitized signals identified by PSD can be used to compute average neutron and $\gamma$-ray signal shapes.
The direct comparison of average pulse shapes from different scintillators only gives partial information on their discrimination capability. In particular this comparison does not consider the effect of the fluctuations discussed above, as opposed to the pulse shape figure of merit $m$.
Nevertheless, average signal shapes can help understand systematic features of PSD.
Fig. \ref{FigAvSignals80nsBC501AStil} shows as examples the neutron and $\gamma$-ray average pulse shapes of stilbene, BC501A and EJ299 in the range [-20, 80] ns, with $t=0$ chosen as the time of the maximum sample.
These average signals were calculated from digitized traces with energy $E_e$ in the range [300, 310] keVee, and their integral was normalised to 1.
The short vertical full line indicates for each scintillator the optimal start time of the slow gate obtained by maximizing the overall figure of merit $M$.
For a given scintillator, one might intuitively expect this slow gate optimal start time to correspond to the time at which the neutron and $\gamma$-ray signals cross.
However, Fig. \ref{FigAvSignals80nsBC501AStil} shows that the optimal start time is larger than the crossing time. This is systematically observed for all the scintillators characterized here.
This is understood easily by noting that the optimal start time of the slow integration in fact maximizes $m$, whereas starting the slow integration when the neutron and $\gamma$-ray signals cross gives only the largest $\overline{D}_n-\overline{D}_\gamma$ difference.
This illustrates the ability of $m$ to correctly account for the effects of pulse shapes on the discrimination quality.
This is also consistent with the decomposition of the overall figure of merit $M$ obtained above, since the $\sqrt{Q}$ factor in $M$ should be insensitive to the start time of the slow integration.

\begin{figure}[htbp]
\begin{center}
\includegraphics[width=7cm]{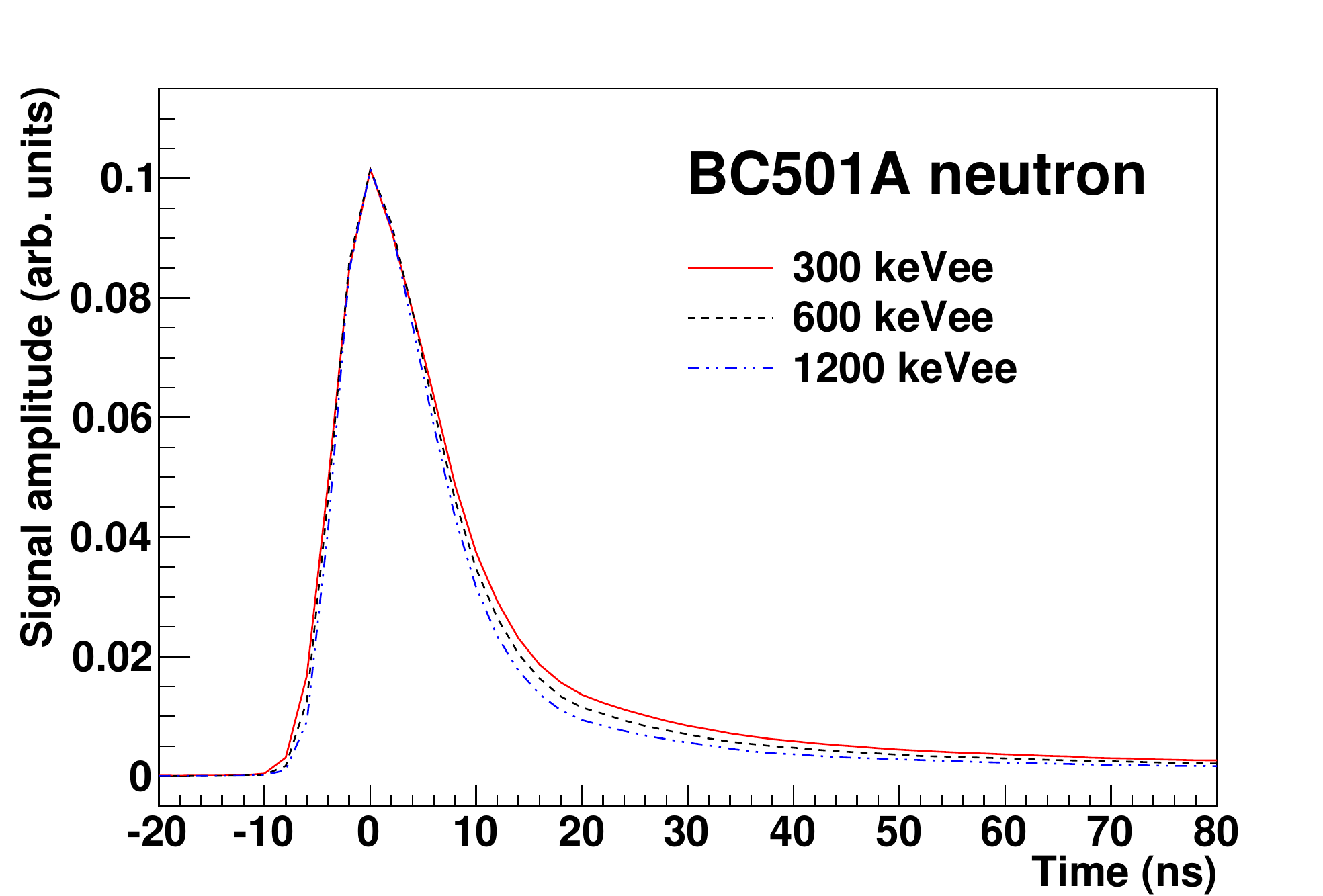}
\includegraphics[width=7cm]{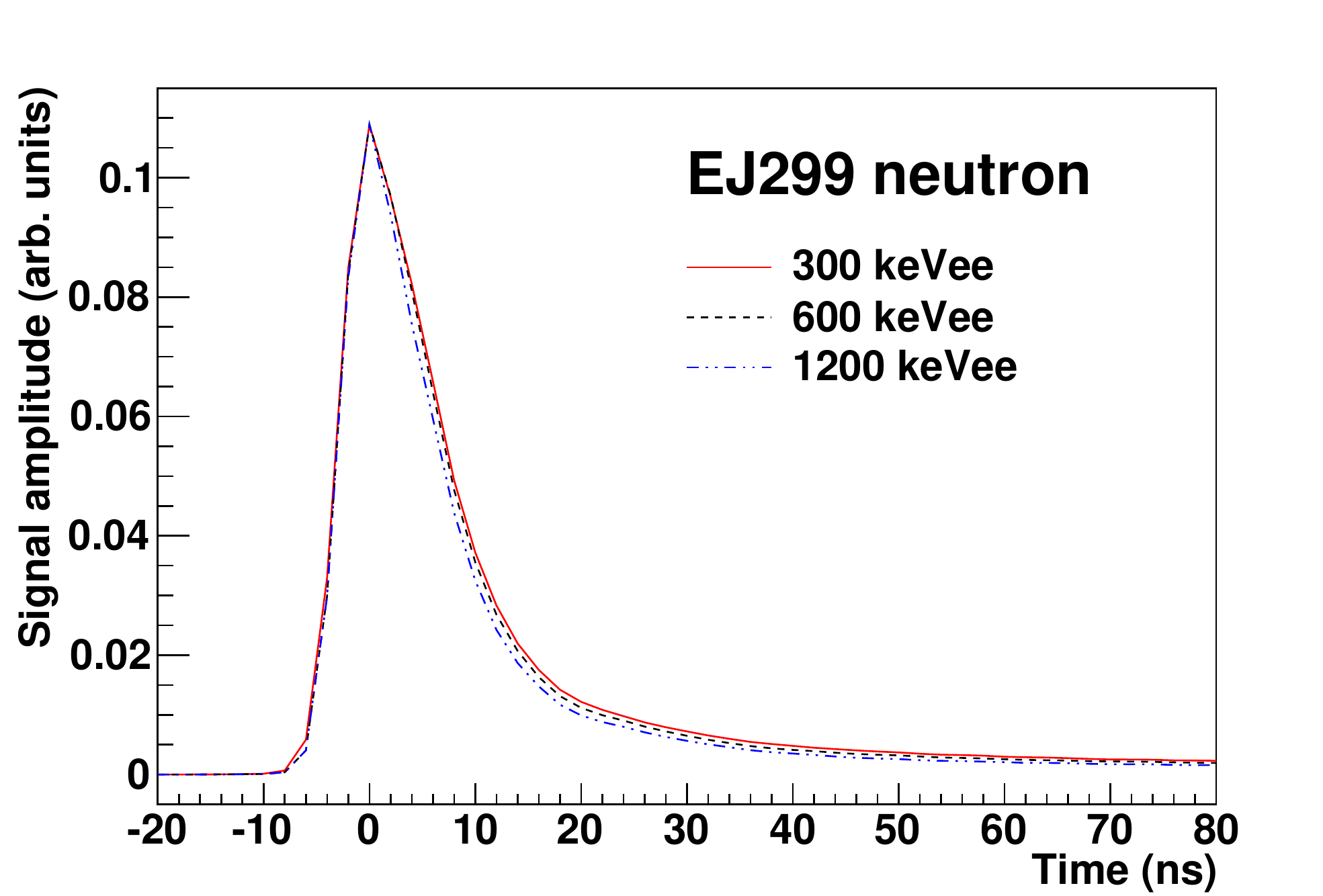}
\caption{Average neutron signals of BC501A and EJ299 at equivalent electron energies of 300, 600 and 1200 keVee (10-keVee wide intervals), normalized to the same pulse height.
The time of the maximum sample was taken as $t=0$. Statistical errors (not shown) are of the order of the thickness of the lines.}
\label{FigAvNeutronSignalsE}
\end{center}
\end{figure}

As noted above, the neutron-induced pulse shape evolves significantly with the energy, while the shape of $\gamma$-ray signals shows a much smaller energy dependence. Examples of neutron average signals at energies of 300, 600 and 1200 keVee (10-keVee wide intervals) are shown on Fig. \ref{FigAvNeutronSignalsE} for BC501A and EJ299.
The amplitude of the signal tail decreases with energy, in agreement with previous works \cite{CesterPSDDig,Guerrero} and with the behaviour of the neutron slow-to-total ratio observed on Fig. \ref{FigHDisc} and Fig. \ref{FigDStopping}.
Interestingly, the energy dependence of the EJ299 neutron pulse shape seems weaker.
We attribute this feature to the smaller sensitivity of the EJ299 slow-to-total ratio to the stopping power, as can be observed on Fig. \ref{FigDStopping}.

\section{Discussion}

\subsection{Limitations of organic scintillators}

It is instructive to estimate the $D_\gamma$ and $D_n$ slow-to-total ratios that can be expected from organic scintillators.
In general, the slow-to-total ratio can be estimated by:
\begin{equation}
D = \frac{\eta_{ionTTA} \, d_{del} + (\eta_{ion}+\eta_{ex}) \, d_{prompt}}{\eta_{ionTTA} + \eta_{ion} + \eta_{ex}},
\end{equation}
with $\eta_{ex}$ the scintillation efficiency for the prompt component from directly excited singlet states, $\eta_{ion}$ the scintillation efficiency for the prompt component from $S_1$ states populated by ionization and recombination, $\eta_{ionTTA}$ the scintillation efficiency for the delayed component from $S_1$ states produced by TTA following ionization and recombination to $T_1$ states, $d_{del}$ ($d_{prompt}$) the fraction of delayed (prompt) component integrated in the slow gate.
By scintillation efficiency we mean the fraction of the energy deposited by the exciting particle that is eventually emitted as scintillation \cite{Birks64}.

For the prompt scintillation from directly excited singlet states, excitation occurs to highly excited $S_n$ states which in the absence of ionization quenching (as \textit{e.g.} for an electron) decay to $S_1$ states with unit efficiency.
The corresponding scintillation efficiency $\eta_{ex}$ can then be estimated by \cite{Birks64}:
\begin{equation} \label{EqScintEff}
\eta_{ex} = F_\pi \, f_{ex} \, \frac{E_0}{E_{ex}} \, q,
\end{equation}
with $F_\pi$ the $\pi$-electron fraction of the scintillator, $f_{ex}$ the fraction of the deposited energy expanded in direct excitations to $S_n$ states of average excitation energy $E_{ex}$, $E_0$ the energy of an $S_1 \rightarrow S_0$ photon and $q$ the $S_1 \rightarrow S_0$ fluorescence quantum yield. This calculation assumes a unitary scintillator (\textit{i.e.} containing a single molecular species), but a similar $D$ would be obtained for a binary scintillator (solvent + solute).

Following ionization, recombination populates $S_n$ and $T_n$ states with respective probabilities $r_S$ and $r_T$.
$S_n$ and $T_n$ states decay to $S_1$ and $T_1$ states, respectively, with unit efficiency in the absence of quenching.
The efficiency for the prompt scintillation from these $S_1$ states is then given by:
\begin{equation}
\eta_{ion} = F_\pi \, (1-f_{ex}) \, r_{S} \, \frac{E_0}{E_{ion}} \, q,
\end{equation}
where $E_{ion}$ is the $\pi$-electron ionization energy.

The scintillation efficiency of delayed fluorescence from $S_1$ states populated by TTA is given by:
\begin{equation}
\eta_{ionTTA} = F_\pi \, (1-f_{ex}) \, r_{T} \, \epsilon_{TTA} \, \frac{E_0}{2 E_{ion}} \, q,
\end{equation}
where $\epsilon_{TTA}$ is the TTA efficiency, \textit{i.e.} the number of $S_1$ states produced per pair of $T_1$ states.

We note that $F_\pi$, $E_0$ and $q$ are involved in all the above scintillation efficiencies, therefore their exact values do not affect the slow-to-total ratio.
For the other parameters, we adopt the following values:
\begin{itemize}
\item $r_{S} = 1/4$ and $r_{T} = 3/4$, as recombinations following $\pi$-electron ionizations populate $T$ and $S$ excited states with a ratio 3:1 given by the ratio of their multiplicities \cite{BrooksReviewNIM}.
\item $d_{prompt} < 5 \%$ for a prompt component consisting of a single exponential decay with a time constant of a few ns.
\item $d_{del} \approx 80$ \% (estimation using parameters from \cite{Voltz}).
\item $E_{0} \approx E_{S_1} \approx 3$ eV, with $E_{S_1}$ the excitation energy of an $S_1$ state.
\item $E_{ex} \approx 1.5 E_{S_1} \approx 4.5$ eV \cite{Birks64}.
\item $E_{ion}\approx$ 8 eV \cite{NIST}.
\item $\epsilon_{TTA} \approx 40$ \% \cite{Birks70}.
\end{itemize}

For electrons from $\gamma$-ray interactions, ionization quenching is negligible and the fraction of energy deposited in direct excitations is $f_{ex} \approx 67$ \% \cite{Birks64}.
With the above parameters, we then estimate a $\gamma$-ray slow-to-total ratio $D_\gamma$ of 4 to 8 \%.
For neutrons that produce heavy recoil particles with high stopping powers, significant ionization quenching occurs.
To estimate $D_n$ we assume that the prompt component from direct excitations is reduced to a negligible intensity by ionization quenching, \textit{i.e.} $\eta_{ex} \approx 0$.
In the case of ionization, recombination to $S_n$ and $T_n$ states is followed by non-radiative $T_n \rightarrow T_1$ and $S_n \rightarrow S_1$ transitions.
Ionization quenching is expected to reduce the efficiency of these transitions. However, as both transition types are allowed transitions, quenching probably affects them equally.
Therefore, the triplet-to-singlet population ratio was assumed to be unaltered by quenching.
In this case, we estimate a neutron slow-to-total ratio $D_n \approx 25-35$ \%.
While measured $\overline{D}_\gamma$ tend to be larger than our rough estimate,
most experimental $\overline{D}_n$ values lie in the estimated range, except for BC537 and CEA-PS where they are smaller than 25 \%.
It is only for NE213-BC501A and stilbene, which give the highest pulse shape figures of merit $m$, that both $\overline{D}_\gamma$ and $\overline{D}_n$ are consistent with our estimates.
This might indicate that these values correspond in fact to the optimum that could be obtained with organic scintillators.

Turning to the light yield, the value for $p$-terphenyl, the maximum for the scintillators considered here,
corresponds to 8 \% of the energy deposited by an electron being reemitted as scintillation light. Although this number may seem small, it is of the order of the limit on the electron scintillation efficiency of pure or binary organic scintillators. Indeed, with the above parameter values and $F_\pi \approx 0.15$ and $q \approx 1$, a total scintillation efficiency $\eta_{ex} + \eta_{ion} + \eta_{ionTTA} \approx 7.4$ \% is estimated for electrons.

Interestingly, stilbene provides the best discrimination with neither the highest light yield nor the largest pulse shape figure of merit $m$.
Among the other scintillators, doped $p$-terphenyl has the largest light yield but a smaller $m$, whereas NE213-BC501A presents the highest $m$ and a smaller light yield.
Therefore, an optimal discrimination seems to result from a compromise between the effects of pulse shapes and light yield.
We note that these characteristics are most likely linked through photophysical properties of the scintillator.
For example, ionization quenching is expected to affect the pulse shapes and the scintillation output in directions that have opposite effects on the PSD quality: a more efficient quenching will tend to increase the delayed light fraction, leading to larger differences between neutron and $\gamma$-ray pulse shapes, while it will decrease the overall scintillation efficiency.
This also suggests that it is impossible to maximize simultaneously the light yield and the discrimination performance.
In binary plastic scintillators,
the high solute concentration required to optimize the discrimination leads 
to interactions between solute molecules that reduce the light yield (``concentration quenching''). The concentration optimal for light yield is therefore smaller than the one optimal for discrimination \cite{Zaitseva}.
Fixing the solute concentration at the optimum for discrimination, the light yield can be increased by adding a secondary solute \cite{Zaitseva}, which in turn produces an increase in discrimination quality.
If the light yield increases from $Y_1$ to $Y_2$, assuming that the pulse shape figure of merit $m$ is unaffected by the secondary solute, on the basis of the light yield dependence of the figure of merit $M$ discussed above one expects an increase of $M$ by a factor $\sqrt{Y_2/Y_1}$. The results reported in Ref. \cite{Zaitseva} reveal that the increase in $M$ is smaller than expected, indicating that the pulse shape figure of merit $m$ is in fact reduced, \textit{i.e.} that the pulse shapes become less favorable for discrimination with the addition of the secondary solute.

\subsection{Alternative scintillating materials}
Given that, as emphasized above, characteristics of some organic scintillators influencing their PSD performance appear to be close to the optimum, such as for stilbene,
significant PSD improvements are expected to require the exploration of new types of materials. The so-called ``triplet harvesting plastic scintillators'', consisting of a polymer matrix (and possibly a scintillating solute, as in a usual binary plastic) doped with a metal complex, have been recently proposed \cite{Feng,ZhmurinPSDPlastic}. In these materials, the prompt scintillation component is emitted by $S_1$ singlet states, either those of the matrix or those of the solute after $S_1$ energy transfer from the matrix. The $T_1$ triplet states of the matrix are efficiently transferred to the metal complex. The high atomic number of the metal ion enhances the spin-orbit interaction responsible for the coupling between $S$ and $T$ states, thus increasing the $T_1 \rightarrow S_0$ radiative transition rate. This emission by the complex gives rise to a delayed component with an exponential decay characterised by a time constant of the order of 1 $\mu$s.
The ratio of the intensities of the delayed and prompt components depends on the nature of the exciting particle, thus allowing PSD.
Furthermore, the spectra of the two scintillation components are different and do not overlap strongly, which could help distinguish the two components through a spectral analysis.
The emission from the complex triplet states does not involve triplet-triplet interactions and is thus more efficient than TTA, leading to larger slow-to-total ratios compared to usual organic scintillators ($\overline{D}_\gamma \approx 40-45$ \% and $\overline{D}_n \approx 50-55$ \% \cite{Feng,ZhmurinPSDPlastic}).
However, since the delayed component is emitted by a unimolecular process, its rate is independent of the triplet density, as shown by its exponential decay, contrary to the bimolecular TTA. This is expected to lead to a weaker sensitivity of pulse shapes to the stopping power of the exciting particle.
This is supported by the fact that the $\gamma$-ray and neutron slow-to-total ratios are very similar, with a relative difference of $\approx$ 15-30 \%, whereas they differ much more in usual scintillators (factor 2 to 4).
Consequently the pulse shape figure of merit $m$ of these materials is limited: with the reported slow-to-total ratios \cite{Feng,ZhmurinPSDPlastic}, we obtain $m \approx 0.1$, a value smaller than for any of the organic scintillators considered here. This seems to set a limit on the PSD performance of these materials smaller than for conventional organic scintillators.

\section{Summary and conclusions}

We have characterized the neutron-$\gamma$ pulse shape discrimination with seven organic scintillators:
the EJ299-33 plastic, a plastic prototype from CEA (CEA-PS), single crystals of stilbene and of the recently introduced doped $p$-terphenyl, the NE213-BC501A liquid and the BC537 deuterated liquid.

We compared the discrimination quality of the different scintillators 
and determined the neutron energies at which the neutron-$\gamma$ discrimination quality reaches a value that can generally be considered as a threshold for good discrimination.
Stilbene was found to offer the best discrimination, followed by doped $p$-terphenyl and by NE213-BC501A.
However, due to a less efficient response to protons, the neutron energy threshold of stilbene results similar to that of doped $p$-terphenyl, around 500 keV.
The neutron threshold obtained with NE213-BC501A is only 10 \% larger.
The EJ299-33 plastic offers a good discrimination over most of the energy range, and can thus be a useful alternative to crystals or to NE213-BC501A. Its neutron energy threshold for good discrimination is determined to be 900 keV.
The BC537 deuterated liquid also gives an efficient discrimination over a large fraction of the energy range, but is more rapidly limited at low energy, hence giving a neutron threshold of 1550 keV.

With simple statistical assumptions, the discrimination figure of merit $M$ was expressed as the product of a ``pulse shape figure of merit'' $m$, measuring the intrinsic contribution of neutron- and $\gamma$-ray-induced signal shapes, and a $\sqrt{Q}$ factor giving the influence of the total light measured by the charge $Q$.
As expected from this factorisation, the experimental $M/m$ ratios, in which pulse shape effects are compensated, follow a $\sqrt{Q}$ trend common to all scintillators studied here. In addition to explaining the overall reduction of the figures of merit as the energy decreases, this $\sqrt{Q}$ term quantifies the effect of the different light yields of the scintillators on their discrimination.
The factorisation of the figure of merit introduced here allows a detailed comparison of the characteristics of scintillating materials. It can also be used for example to evaluate the effect of a modification of light output and/or pulse shapes of these materials and as such can guide the search for improved formulations. As illustrated on conventional organic scintillators and modified plastic scintillators, it also provides a means to quantify the intrinsic limitations of the different materials.

The measurements performed here show that stilbene gives the best PSD although it has neither the highest light yield nor the highest pulse shape figure of merit. In comparison, NE213-BC501A, with the most favourable pulse shapes for PSD but a lower light yield, and doped $p$-terphenyl, with the highest light yield but a much smaller pulse shape figure of merit, provide inferior PSD performance. Good discrimination thus seems to result from a compromise between pulse shapes and scintillation output. In that respect, stilbene might be close to the optimal that can be obtained with organic scintillators.
Compared to stilbene, PSD with the EJ299-33 plastic and CEA-PS plastic prototype seems to be limited by both a lower light yield and less beneficial pulse shapes. This is also the case for the BC537 deuterated liquid scintillator.

Alternative plastic scintillators, containing a metal complex that enhances the direct fluorescence emission from triplet states, present a larger fraction of delayed scintillation than in usual organic scintillators. However, the lower pulse shape figure of merit of these materials indicates that their scintillation pulse shapes are intrinsically less favourable for PSD compared to those from conventional organic scintillators.

\section*{Acknowledgments}

The authors are grateful to Dr M. Hamel for kindly providing the CEA plastic scintillator sample.

\onecolumn

\begin{table}[htbp]
\begin{center}
\begin{tabular}{lccc}
\hline
Scintillator & $\lambda_{max}$ & {Light yield} & Measured light yield \\
  & (nm) & Ph./MeVee & $Y/Y_\text{BC501A}$ \\
\hline
$p$-terphenyl  &  420  &  27000   & 2.18(11) \\
Stilbene        &  390  &  14000   & 1.88(10) \\
BC501A       &  425  &  12000   & \\
NE213         &  425  &  12000   & 1.02(5) \\
BC537         &  425  &    9200   &  0.80(4) \\
EJ299          &  420  &    8600   & 1.31(7)  \\
CEA-PS         &   -      &     -         & 0.58(3) \\
\hline
\end{tabular}
\caption{Wavelength of maximum emission $\lambda_{max}$, light yield in photons for 1 MeV deposited by electrons (Ph./MeVee) from manufacturer data, and measured light yields relative to BC501A $Y/Y_\text{BC501A}$.}
\label{TabYields}
\end{center}
\end{table}

\begin{table}[htbp]
\begin{center}
\begin{tabular}{lcccc}
\hline
\\ [-1.9ex]
Scintillator & Slow gate & $E_e^{M=1}$ & $E_n^{M=1}$ & Light output \\
& start time (ns) & (keVee) & (keV) & function \\
\\ [-1.9ex]
\hline
$p$-terphenyl & 20 &    70(4) &   492(22) & \textit{a} \\
Stilbene       & 28 &    57(4) &   517(25) & \textit{b} \\
BC501A      & 20 &     80(5) &  551(21) & \textit{c} \\
NE213        & 20 &     79(5) &  547(22) & \textit{c} \\
BC537        & 16 &   240(7) & 1550(30) & \textit{d} \\
EJ299         & 32 &  100(7) &   900(40) & \textit{e} \\
CEA-PS        & 54 & 990(50) & (3000) & \textit{f} \\
\hline
\end{tabular} \\
\textit{a} \cite{SardetPTerphenyl},
\textit{b} \cite{HansenRichterStilbene},
\textit{c} NE213 function \cite{Cecil},
\textit{d} \cite{CroftNE230},
\textit{e} \cite{LawrenceEJ299},
\textit{f} NE102 function \cite{Cecil}
\caption{Optimal slow gate start times, given with respect to the time of the maximum sample; equivalent electron energy $E_e^{M=1}$ for a figure of merit $M=1$ and corresponding neutron energy $E_n^{M=1}$. Proton (deuteron for BC537) light output functions used to compute the neutron energies are indicated. For BC537, the neutron energy was taken as 9/8 of the deuteron recoil energy.}
\label{TabGatesEM1}
\end{center}
\end{table}

\begin{table}[htbp]
\begin{center}
\begin{tabular}{lccccccc}
\hline
\\ [-1.9ex]
Scintillator & $\overline{D}_\gamma$ & $\overline{D}_n$ & $M$ & $m$ & $\overline{\left({\frac{dE}{\rho dx}}\right)}_e$ & $E_{r}$ & $\overline{\left({\frac{dE}{\rho dx}}\right)}_r$\\

& (\%) & (\%) & & & (MeV cm$^2$/g) & (MeV) & (MeV cm$^2$/g) \\
\\ [-1.9ex]
\hline
$p$-terphenyl  & 11.1 & 30.7 & 2.48(7) & 0.252 & 4.4 & 1.46 & 340 \\
Stilbene        & 8.9 & 32.3 & 2.71(6) & 0.311 & 4.3 & 1.61 & 310 \\
BC501A       & 8.3 & 34.0 & 2.29(7) & 0.343 & 4.8 & 1.28 & 380 \\
NE213         & 8.8 & 34.7 & 2.27(5) & 0.342 & 4.8 & 1.28 & 380 \\
BC537         & 6.3 & 20.3 & 1.17(2) & 0.217 & 3.8 & 1.61 & 470 \\
EJ299          & 12.8 & 29.8 & 1.64(4) & 0.214 & 4.7 & 1.79 & 310 \\
CEA-PS         & 8.1 & 18.3 & 0.70(5) & 0.155 & 4.5 & 1.35 & 360 \\
\hline
\end{tabular}
\caption{Quantities characterizing the discrimination at 300 keVee: $\gamma$-ray and neutron mean slow-to-total ratios $\overline{D}_\gamma$ and $\overline{D}_n$, figure of merit $M$, pulse shape figure of merit $m$, average mass stopping power $\overline{\left({\frac{dE}{\rho dx}}\right)}_e$ for an electron, energy $E_{r}$ of a neutron scattering recoil particle (proton, or deuteron for BC537) and average mass stopping power $\overline{\left({\frac{dE}{\rho dx}}\right)}_r$ for the recoil particle.
Absolute errors on $\overline{D}_\gamma$ and $\overline{D}_n$ are smaller than 0.03 and 0.05 \% respectively (0.13 and 0.14 for CEA-PS respectively), and errors on $m$ are smaller than $2\times 10^{-3}$ ($5\times 10^{-3}$ for the CEA-PS).}
\label{TabPSD300keV}
\end{center}
\end{table}

\begin{thebibliography}{50}

\bibitem{ORION} E. Li\'enard et al., Nucl. Instr. and Meth. in Phys. Res. A 413 (1998) 321.

\bibitem{BNB} U. Jahnke et al., Nucl. Instr. and Meth. in Phys. Res. A 508 (2003) 295.

\bibitem{CARMEN} X. Ledoux et al., Nucl. Instr. and Meth. in Phys. Res. A 844 (2017) 24.

\bibitem{EDEN} H. Laurent et al., Nucl. Instr. and Meth. in Phys. Res. A 326 (1993) 517.

\bibitem{DEMON} I. Tilquin et al., Nucl. Instr. and Meth. in Phys. Res. A 365 (1995) 446.

\bibitem{NWALL} P. D. Zecher et al., Nucl. Instr. and Meth. in Phys. Res. A 401 (1997) 329.

\bibitem{TONNERRE} A. Bu\c{t}\u{a} et al., Nucl. Instr. and Meth. in Phys. Res. A 455 (2000) 412.

\bibitem{Normand} S. Normand et al., Nucl. Instr. and Meth. in Phys. Res. A 484 (2002) 342.

\bibitem{Bass} C. D. Bass et al., Appl. Radiat. Isot. 77 (2013) 130.

\bibitem{Birks64} J.B. Birks, The Theory and Practice of Scintillation Counting, Pergamon Press, London, 1964.

\bibitem{Knoll} G.F. Knoll, Radiation Detection and Measurements, third ed., John Wiley and
Sons, New York, 2000 (and references therein).

\bibitem{BrooksReviewNIM} F. D. Brooks, Nucl. Instr. and Meth. 162 (1979) 477.

\bibitem{Brooks1959} F. D. Brooks, Nucl. Instr. and Meth. 4 (1959) 151.

\bibitem{Brooks1974} F. D. Brooks and D. T. L. Jones, Nucl. Instr. and Meth. 121 (1974) 69.

\bibitem{Budakovsky} S. Budakovsky et al., IEEE Trans. Nucl. Sci. NS54 (2007) 2734.

\bibitem{Brooks60} F.D. Brooks, R.W. Pringle, B.L. Funt, I.R.E.Trans. Nucl. Sci. NS-7 (2Ð3) (1960) 35.

\bibitem{Winyard} R. A. Winyard, J. E. Lutkin and G. W. McBeth, Nucl. Instr. and Meth., 95 (1971) 141.

\bibitem{Zaitseva} N. Zaitseva et al., Nucl. Instr. and Meth. in Phys. Res. A 668 (2012) 88.

\bibitem{Eljen} Eljen Technology, www.eljentechnology.com.

\bibitem{Zaitseva2018} N. P. Zaitseva et al., Nucl. Instr. and Meth. in Phys. Res. A 889 (2018) 97.

\bibitem{Blanc} P. Blanc et al., Nucl. Instr. and Meth. in Phys. Res. A 750 (2014) 1.

\bibitem{Zhmurin} P. N. Zhmurin et al., Nucl. Instr. and Meth. in Phys. Res. A 761 (2014) 92.

\bibitem{PozziEJ299} S. A. Pozzi, M. M. Bourne and S. D. Clarke, Nucl. Instr. and Meth. in Phys. Res. A 723 (2013) 19.

\bibitem{NyibuleEJ299} S. Nyibule et al., Nucl. Instr. and Meth. in Phys. Res. A 728 (2013) 36.

\bibitem{CesterEJ299} D. Cester et al., Nucl. Instr. and Meth. in Phys. Res. A 735 (2014) 202.

\bibitem{LawrenceEJ299} C. C. Lawrence et al., Nucl. Instr. and Meth. in Phys. Res. A 759 (2014) 16.

\bibitem{CryosBeta} Cryos-Beta, cryos-beta.kharkov.ua.

\bibitem{Moszynski} M. Moszynski et al., Nucl. Instr. and Meth. in Phys. Res. A 350 (1994) 226.

\bibitem{SaintGobain} Saint-Gobain Crystals, www.crystals.saint-gobain.com.

\bibitem{FASTER} http://faster.in2p3.fr/.

\bibitem{TheseSenoville} M. S\'enoville, Th\`ese de doctorat (PhD thesis), Universit\'e de Caen, 2013, https://tel.archives-ouvertes.fr/tel-01064554v1.

\bibitem{Flynn} K. F. Flynn et al., Nucl. Instr. and Meth. 27 (1964) 13.

\bibitem{Horrocks} D. L. Horrocks, Nucl. Instr. and Meth. 30 (1964) 157.

\bibitem{Dietze} G. Dietze, IEEE Trans. Nucl. Sci. NS-26 (1979) 398.

\bibitem{WoolfEJ299} R. S. Woolf et al., Nucl. Instr. and Meth. in Phys. Res. A 784 (2015) 80.

\bibitem{Heltsley} J. H. Heltsley et al., Nucl. Instr. and Meth. in Phys. Res. A 263 (1988) 441.

\bibitem{MoszynskiLargeVolume} M. Moszynski et al., Nucl. Instr. and Meth. in Phys. Res. A 317 (1992) 262.

\bibitem{CesterPSDDig} D. Cester et al., Nucl. Instr. and Meth. in Phys. Res. A 748 (2014) 33.

\bibitem{Guerrero} C. Guerrero et al., Nucl. Instr. and Meth. in Phys. Res. A 597 (2008) 212.

\bibitem{AsymGaussians} T. S. Buys and K. de Clerk, Anal. Chem., 44, 1273 (1972).

\bibitem{SardetPTerphenyl} A. Sardet et al., Nucl. Instr. and Meth. in Phys. Res. A 792 (2015) 74.

\bibitem{HansenRichterStilbene} W. Hansen and D. Richter, Nucl. Instr. and Meth. in Phys. Res. A 476 (2002) 195.

\bibitem{Cecil} R. A. Cecil, B. D. Anderson and R. Madey, Nucl. Instr. and Meth. 161 (1979) 439.

\bibitem{CroftNE230} S. Croft et al., Nucl. Instr. and Meth. in Phys. Res. A 316 (1992) 324.

\bibitem{Voltz} R. Voltz, H. Dupont and G. Laustriat, J. Physique 29 (1968) 297.

\bibitem{NIST} NIST Chemistry WebBook, www.webbook.nist.gov.

\bibitem{Birks70} J. B. Birks, Chem. Phys. Lett. 7 (1970) 293.

\bibitem{Feng} P. L. Feng et al., IEEE Trans. Nucl. Sci. NS59 (2012) 3312.

\bibitem{ZhmurinPSDPlastic} P. N. Zhmurin et al., Rad. Meas. 62 (2014) 1.

\end{thebibliography}
\end{document}